\documentclass[fleqn,useAMS,usenatbib,usegraphicx]{mn2e}
\usepackage{graphicx}
\usepackage{rotating}
\usepackage{float}
\usepackage{times}
\usepackage{natbib}
\usepackage{epsfig}
\usepackage{lscape}
\usepackage{amssymb}

\newcommand{\kms}{km\,s$^{-1}$}

\newcommand{\hi}{H{\sc i}}
\newcommand{\hii}{H{\sc i}~21\,cm}

\begin{document}

\title[H{\sc i}~21\,cm absorption in high-$z$ flat spectrum sources]
{A Giant Metrewave Radio Telescope search for associated H{\sc i}~21\,cm absorption in high-redshift flat-spectrum 
sources}

\author[Aditya et al.]{J.~N.~H.~S. Aditya\thanks{E-mail: aditya@ncra.tifr.res.in (JNHSA); nkanekar@ncra.tifr.res.in (NK); sushma@ncra.tifr.res.in (SK)}, 
Nissim~Kanekar\thanks{Swarnajayanti Fellow}, Sushma Kurapati\\
National Centre for Radio Astrophysics, Tata Institute of Fundamental Research, Pune 411007, India
}


\maketitle

\begin{abstract}
We report results from a Giant Metrewave Radio Telescope search for ``associated'' redshifted H{\sc i}~21\,cm 
absorption from 24 active galactic nuclei (AGNs), at $1.1 < z < 3.6$, selected from the Caltech-Jodrell Bank 
Flat-spectrum (CJF) sample. 22 out of 23 sources with usable data showed no evidence of absorption, with 
typical $3\sigma$ optical depth detection limits of $\approx 0.01$ at a velocity resolution of 
$\approx 30$~km~s$^{-1}$. A single tentative absorption detection was obtained at $z \approx 3.530$ 
towards TXS\,0604+728. If confirmed, this would be the highest redshift at which H{\sc i}~21\,cm absorption 
has ever been detected. 

Including 29 CJF sources with searches for redshifted H{\sc i}~21\,cm absorption in the literature, mostly at 
$z < 1$, we construct a sample of 52 uniformly-selected flat-spectrum sources. A Peto-Prentice two-sample test 
for censored data finds (at $\approx 3\sigma$ significance) that the strength of H{\sc i}~21\,cm absorption is 
weaker in the high-$z$ sample than in the low-$z$ sample; this is the first statistically significant evidence for 
redshift evolution in the strength of H{\sc i}~21\,cm absorption in a uniformly selected AGN sample. However, the 
two-sample test also finds that the H{\sc i}~21\,cm absorption strength is higher in AGNs with 
low ultraviolet or radio luminosities, at $\approx 3.4 \sigma$ significance. The fact that the higher-luminosity 
AGNs of the sample typically lie at high redshifts implies that it is currently not possible to break the 
degeneracy between AGN luminosity and redshift evolution as the primary cause of the low H{\sc i}~21\,cm 
opacities in high-redshift, high-luminosity active galactic nuclei.

\end{abstract}


\section{Introduction} 
The generally accepted model of an active galactic nucleus (AGN) consists of a central super-massive 
black hole surrounded by an accretion disk and clouds of ionized, atomic and molecular gas. The 
energy output from the AGN is thought to be fuelled by the supply of gas onto the central black hole 
\citep[e.g.][]{rees84}. The gas may be supplied to the central regions 
either through slow accretion of material from larger scales, or through the triggering of gas infall 
by galaxy mergers \citep[e.g.][]{struve10}. Conversely, the AGN also may give rise to energetic gas 
outflows, resulting in the quenching of star formation in the central regions, and possibly even 
ending the active state of the nucleus \citep[e.g.][]{fabian12}. Studies of the distribution 
and kinematics of the circumnuclear gas allow us to probe the relative importance of infall and outflows 
at different redshifts, yielding insights into the fuelling 
of AGN activity and feedback mechanisms, both of which have strong influence on galaxy evolution.

In the case of radio-loud AGNs, absorption studies in the \hii\ line provide an interesting probe of 
gas kinematics in the nuclear regions \citep[see, e.g.,][for a recent review]{morganti12}. Such 
``associated'' \hii\ absorption studies can be used to test AGN unification schemes 
\citep[e.g.][]{barthel89}.  These scenarios predict that the line of sight to broad-line AGNs
is normal to the torus (and to the thick disk), while that toward narrow-line AGNs lies close to the 
plane of the torus. Unification schemes hence suggest that associated \hii\ absorption should be 
systematically more common in the latter class of systems, as the nucleus in broad-line systems is 
not expected to be obscured by the neutral gas. Associated \hii\ absorption studies can also be used 
to identify and study circumnuclear disks and tori around AGNs \citep[e.g.][]{conway95,carilli98c,peck02,morganti08}. 
Finally, associated absorbers with high \hi\ column densities are also likely to show molecular absorption 
in transitions of species like CO, HCO$^+$, OH, etc. 
\citep[e.g.][]{carilli92,carilli97,wiklind94,wiklind96,kanekar02,kanekar08c}. Such systems allow 
further detailed studies of physical and chemical properties of the AGN environment. 
In addition, comparisons between the redshifts of different atomic and molecular transitions in such systems 
allow one to probe the possibility of changes in the fundamental constants of physics, such as the fine structure
constant and the proton-electron mass ratio \citep[e.g.][]{drinkwater98,carilli00,chengalur03,kanekar04b,kanekar10b}.

While searches for associated \hii\ absorption have been carried out for about three decades, 
conditions in the neutral gas in AGN environments are still unclear.  The first such study was 
that of \citet{vangorkom89} who used the Very Large Array to search for \hii\ absorption in a well-defined 
sample of nearby radio galaxies. They found that the detected \hii\ absorption features tend to be redshifted 
from the systemic velocity of the targets, i.e. that gas infall appears to be more common than gas outflow 
in local AGNs. Interestingly, their estimated mass supply rate from the gas velocities was much larger than 
required to fuel the AGN radio activity (only $\approx 10^{-5} - 10^{-3}$~M$_\odot$~yr$^{-1}$), indicating that cold atomic 
gas might be the main fuel source for this activity. Later, however, \citet{vermeulen03} used the Westerbork 
Synthesis Radio Telescope to search for associated \hii\ absorption in a large, heterogeneous AGN sample at 
intermediate redshifts ($0.2 \lesssim z \lesssim 0.8$) and found comparable numbers of redshifted and 
blueshifted absorption features \citep[see also][]{pihlstrom03}. Their results indicate that both infall 
and outflow may be important in AGN environments, with the complex absorption profiles possibly 
arising from interactions of the neutral gas with the radio jets or gas rotation around the central 
nucleus. Later searches for \hii\ absorption in large, but again heterogeneous, samples 
\citep[e.g.][]{gupta06a,curran08,gereb15} have been unable to shed much light on whether or 
not the inflow of neutral gas is likely to be the main source of AGN fuel. We note, in passing, that 
a strong correlation has recently been found between the accretion rates of hot X-ray emitting gas, 
associated with the hot corona, and the AGN jet power \citep[e.g.][]{allen06}. These results suggest 
that hot X-ray emitting gas could be an important component of the fuel that powers the active nucleus.

The above studies of associated \hii\ absorbers have found absorption features having a wide range 
of line profiles (e.g. both narrow and broad), and differing offsets from the systemic velocity, 
indicating complex morphologies of the gaseous structures around the central core 
\citep[e.g.][]{gupta06a, gereb15}. Narrow lines with smaller velocity offsets ($\leq 100$~\kms) from 
the AGN redshift would indicate gas clouds with small velocity dispersions, possibly rotating in 
circumnuclear disks, whereas broader absorption features suggest unsettled gas, perhaps interacting 
with the central radio source. Features with the largest widths ($\gtrsim 200$~\kms) are likely to 
arise due to interactions between the gas and the AGN radio jet. These wide and typically blueshifted
absorption features are more commonly found in compact sources \citep[e.g.][]{pihlstrom03,gupta06a,gereb15}, 
indicating that such sources may be the best targets for searches for radio source-gas interactions 
as well as strong AGN-driven outflows. The outflowing gas is expected to be deposited in the intergalactic 
medium (IGM), transferring energy back to the IGM, and thus regulating gas collapse and hence star 
formation in the AGN host galaxy via feedback. Such feedback mechanisms have been shown to be critical 
in understanding galaxy formation: specifically, simulations of galaxy evolution which do not implement 
feedback models produce far higher star formation rates (SFRs) than the typical SFRs observed in the Universe, 
while those that include some prescription for feedback tend to be consistent with the observed data 
\citep[e.g.][]{springel05}. Despite its acknowledged importance, our understanding of feedback effects in 
galaxies is very limited, and it is as yet even unclear which of the two most prominent feedback 
mechanisms, AGN-induced feedback \citep[e.g.][]{fabian12,morganti13} and supernovae outbursts 
\citep[e.g.][]{efstathiou00}, is the dominant process for the regulation of star formation.
The detection of blueshifted absorption features is a signature of gas outflows and can thus 
be used to trace the importance of feedback at different redshifts, and in different AGN types. 
Follow-up Very Long Baseline Interferometry (VLBI) mapping studies in the redshifted \hii\ line can then 
be possibly used to trace the location of the gas outflows and determine whether they arise from the AGN 
or from supernovae \citep[e.g.][]{morganti13}. 

\hii\ absorption studies thus provide an important probe of the distribution and kinematics of 
neutral hydrogen gas in AGN environments, and their redshift evolution \citep[e.g.][]{kanekar04}.
Unfortunately, previous surveys for associated \hii\ absorption have been mostly limited to low 
redshifts ($z \leq 1$) \citep[e.g.][]{morganti01,vermeulen03,gupta06a}, with only a handful of 
such searches at $z > 1$ in the literature \citep[e.g.][]{gupta06a, curran13}, some of which have 
low sensitivity. Indeed, there are currently only four detections of associated \hii\ absorbers 
at $z \ge 1$, at $z \approx 1.2$ towards 3C190 \citep[][]{ishwar03}, $z \approx 1.3$ towards J1545+4751 
\citep[][]{curran13}, $z \approx 2.6$ towards MG\,J0414+0534 \citep[][]{moore99}, and $z \approx 3.4$ 
towards TXS\,0902+343 \citep[][]{uson91}. 

The main hindrance to using \hii\ absorption studies to probe the redshift evolution of AGN 
environments is the  paucity of \hii\ absorbers at high redshifts, $z > 1$. Further, most studies 
in the literature at higher redshifts \citep[e.g.][]{gupta06a,curran13} have targetted 
heterogeneous samples for which it is often difficult to separate redshift effects from source 
characteristics. We have hence begun a project to use the Giant Metrewave Radio Telescope (GMRT) 
to carry out a search for associated \hii\ absorption in a large sample of high-$z$ flat spectrum 
sources, selected in a uniform manner. In this paper, we present initial results from this project, 
from a GMRT search for redshifted \hii\ absorption in 24 AGNs of the sample, at $z \approx 1.3$ and 
$z \approx 3.5$.

\begin{table*}
\caption{\label{table:data} The 24 targets of this paper, selected from the CJF sample, 
in order of increasing redshift. 
Note that L$'_{\rm UV}=$~Log[L$_{\rm UV}$/W~Hz$^{-1}$] and L$'_{\rm 1.4\,GHz} =$~Log[L$_{\rm 1.4\,GHz}$/W~Hz$^{-1}$].}
\begin{center}
\begin{tabular}{|c|c|c|c|c|c|c|c|c|c|c|}
\hline
Source    & $z$ & $\nu_{\rm 21\,cm}$ & BW  & $\Delta v$ & S$_\nu$ & $\Delta S$ & $\int \tau d{\rm V}$ & L$'_{\rm UV}$ &  L$'_{\rm 1.4\,GHz}$ & $\alpha_{\rm 21\,cm}$                \\
          &     & MHz                & MHz & \kms\      & mJy     & mJy &  \kms\   &  &  & \\
\hline
TXS\,0600+442  & 1.136 & 664.98 & 8.0     & 28.2     & $1260.7 \pm 0.6$ &  4.9  & $<0.71$         & $-$   & 27.61 & -0.37   \\        
TXS\,2356+390  & 1.198 & 646.23 & 8.0$^b$ & 29.0     & $643.0  \pm 0.5$ &  1.2  & $<0.42$         & 22.36 & 27.36 & -1.16   \\     
TXS\,0821+394  & 1.216 & 641.08 & 8.0     & 29.2     & $2532.8 \pm 0.4$ &  5.1  & $<0.44$         & 23.56 & 27.97 & -0.75   \\   
TXS\,0945+408  & 1.249 & 631.56 & 8.0     & 29.7     & $2026.0 \pm 0.5$ &  2.8  & $<0.30$         & 23.93 & 27.89 & -0.46   \\    
TXS\,0641+392  & 1.266 & 626.83 & 8.0     & 29.9     & $417.7  \pm 0.3$ &  1.5  & $<0.68$         & 22.04 & 27.22 &  0.79   \\      
TXS\,0537+531  & 1.275 & 624.35 & 8.0     & 30.0     & $711.3  \pm 0.5$ &  1.2  & $<0.32$         & 22.85 & 27.46 &  0.01   \\      
TXS\,0707+476  & 1.292 & 619.72 & 33.3    & 31.5$^a$ & $1022.5 \pm 0.2$ &  1.4  & $<0.30$         & 23.86 & 27.63 & -0.15   \\   
TXS\,0850+581  & 1.318 & 612.89 & 8.0     & 30.6     & $998.6  \pm 0.5$ &  2.0  & $<0.37$         & 23.53 & 27.63 & -0.26   \\    
S5\,2353+81    & 1.344 & 605.98 & 8.0     & 30.9     & $581.6  \pm 0.6$ &  2.7  & $<0.87$         & 22.31 & 27.41 & -0.53   \\     
TXS\,0035+413  & 1.353 & 603.66 & 8.0     & 31.0     & $541.8  \pm 0.6$ &  2.1  & $<0.76$         & 23.12 & 27.39 &  0.35   \\     
TXS\,1030+611  & 1.401 & 591.60 & 8.0     & 31.7     & $692.1  \pm 0.5$ &  3.1  & $<0.87$         & 23.57 & 27.53 & -0.37   \\     
TXS\,0820+560  & 1.418 & 587.34 & 8.0     & 31.9     & $1507.7 \pm 0.1$ &  2.5  & $<0.34$         & 23.69 & 27.87 & -0.26   \\    
TXS\,0805+410  & 1.418 & 587.32 & 8.0     & 31.9     & $512.3  \pm 0.8$ &  2.0  & $<0.69$         & 23.32 & 27.40 & -0.07   \\     
TXS\,0804+499  & 1.436 & 583.08 & 8.0     & 32.1     & $682.6  \pm 0.5$ &  3.6  & $<0.95$         & 23.45 & 27.54 &  0.31   \\    
TXS\,0917+624  & 1.446 & 580.71 & 8.0     & 32.3     & $975.6  \pm 0.5$ &  4.2  & $<0.79$         & 23.22 & 27.70 &  0.23   \\    
TXS\,0859+470  & 1.470 & 575.05 & 8.0     & 32.6     & $2819.5 \pm 0.4$ &  6.2  & $<0.43$         & 23.68 & 28.17 & -0.23   \\   
TXS\,2253+417  & 1.476 & 573.67 & 8.0     & 32.6     & $1255.2 \pm 0.5$ &  4.5  & $<0.71$         & $-$   & 27.82 &  0.30   \\         
TXS\,0340+362  & 1.484 & 571.82 & 8.0     & 33.0     & $291.4  \pm 0.5$ &  4.7  & $<3.00$         & 21.85 & 27.19 &  0.35   \\      
TXS\,0800+618  & 3.033 & 352.20 & 4.0     & 26.6     & $858.5  \pm 0.5$ &  6.7  & $<1.17$         & 23.67 & 28.22 & -0.05   \\    
S5\,0014+81    & 3.366 & 325.33 & 4.0$^b$ & 28.8     & $773.5  \pm 0.4$ &  RFI  & RFI             & 25.25 & 28.27 & -0.11   \\   
TXS\,0642+449  & 3.396 & 323.11 & 4.0     & 29.0     & $310.6  \pm 1.1$ &  4.2  & $<2.86$         & 24.48 & 27.86 &  0.72   \\    
TXS\,0620+389  & 3.469 & 317.84 & 4.0     & 29.5     & $1608.3 \pm 0.8$ &  8.3  & $<0.20$         & 23.95 & 28.50 & -0.16   \\  
TXS\,0604+728  & 3.530 & 313.56 & 4.0     & 29.9     & $1890.9 \pm 1.9$ &  3.5  & $4.29 \pm 0.28$ & 23.63 & 28.64 & -0.38   \\
TXS\,0749+426  & 3.589 & 309.51 & 4.0     & 30.3     & $537.5  \pm 0.6$ &  2.0  & $<0.76$         & 24.61 & 28.14 &  0.16   \\  
\hline
\end{tabular}
\end{center}
$^a$In the case of TXS\,0707+476, the listed velocity resolution in Column~(5) is without Hanning-smoothing and re-sampling.\\
$^b$The listed bandwidths and resolutions for TXS\,2356+390 and S5\,0014+81 are for the original 
observations with the hardware backend. Both sources were re-observed using the software backend,
with 512 channels; the bandwidth and velocity resolution were 33.3~MHz and $\approx 30.2$~\kms\
for TXS\,2356+390, and 16.7~MHz and $\approx 30.0$~\kms\ for S5\,0014+81 (for both sources, the 
quoted resolutions are without Hanning-smoothing and re-sampling).
\end{table*}


\section{The sample, observations, data analysis and results}

\begin{figure*}
\centering
\begin{tabular}{ccc}

\includegraphics[scale=0.25]{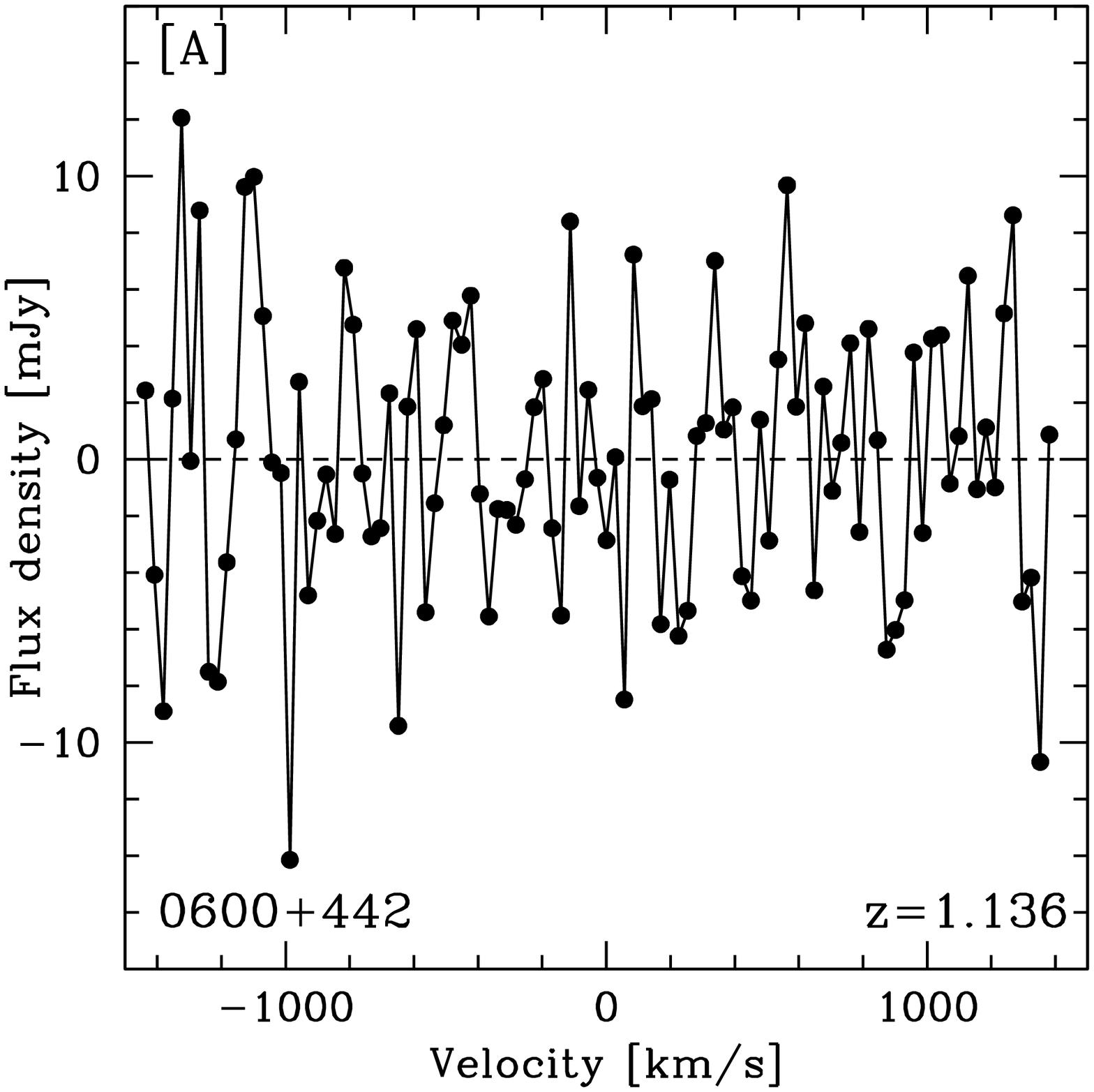} & \includegraphics[scale=0.25]{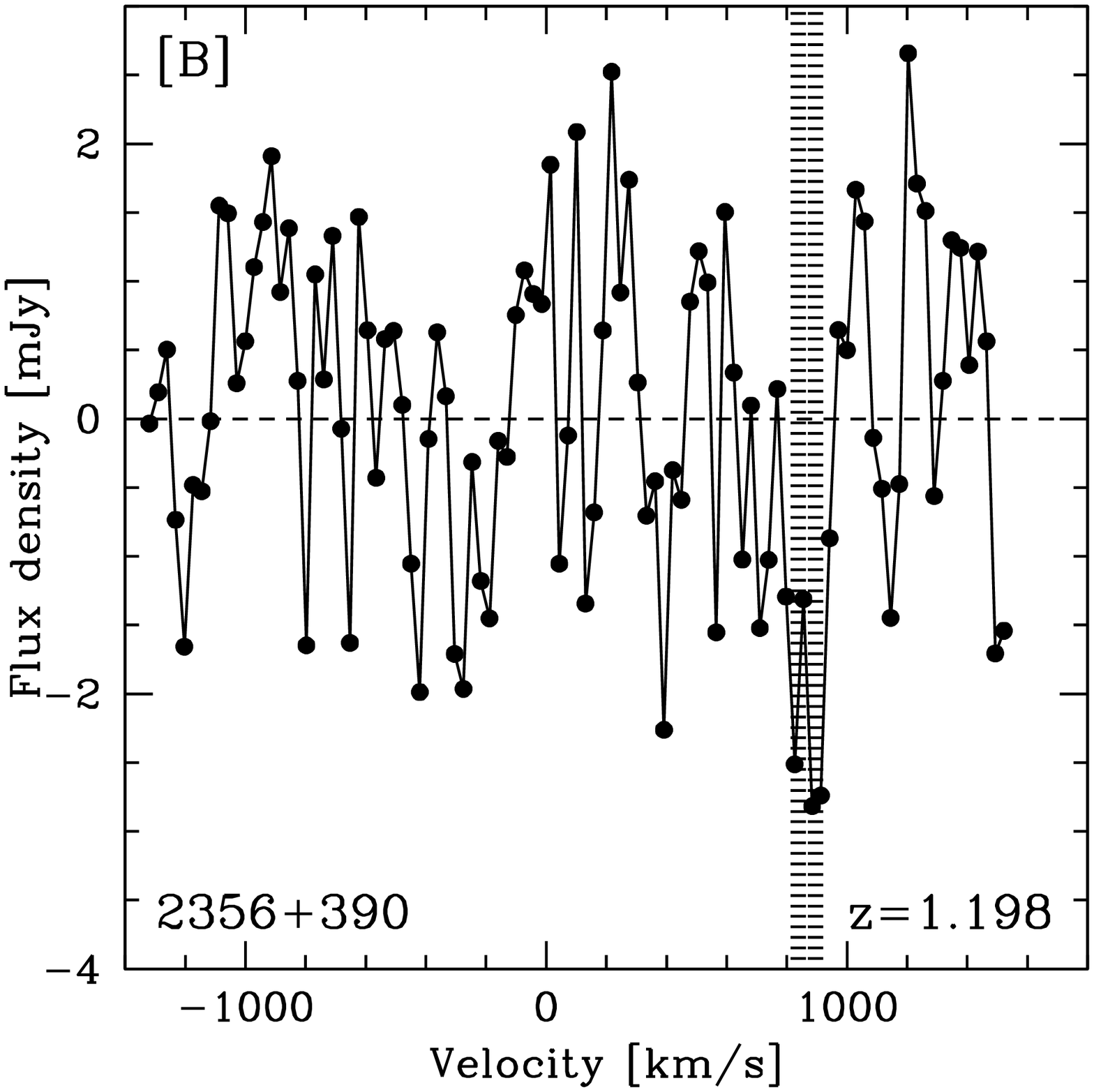} & \includegraphics[scale=0.25]{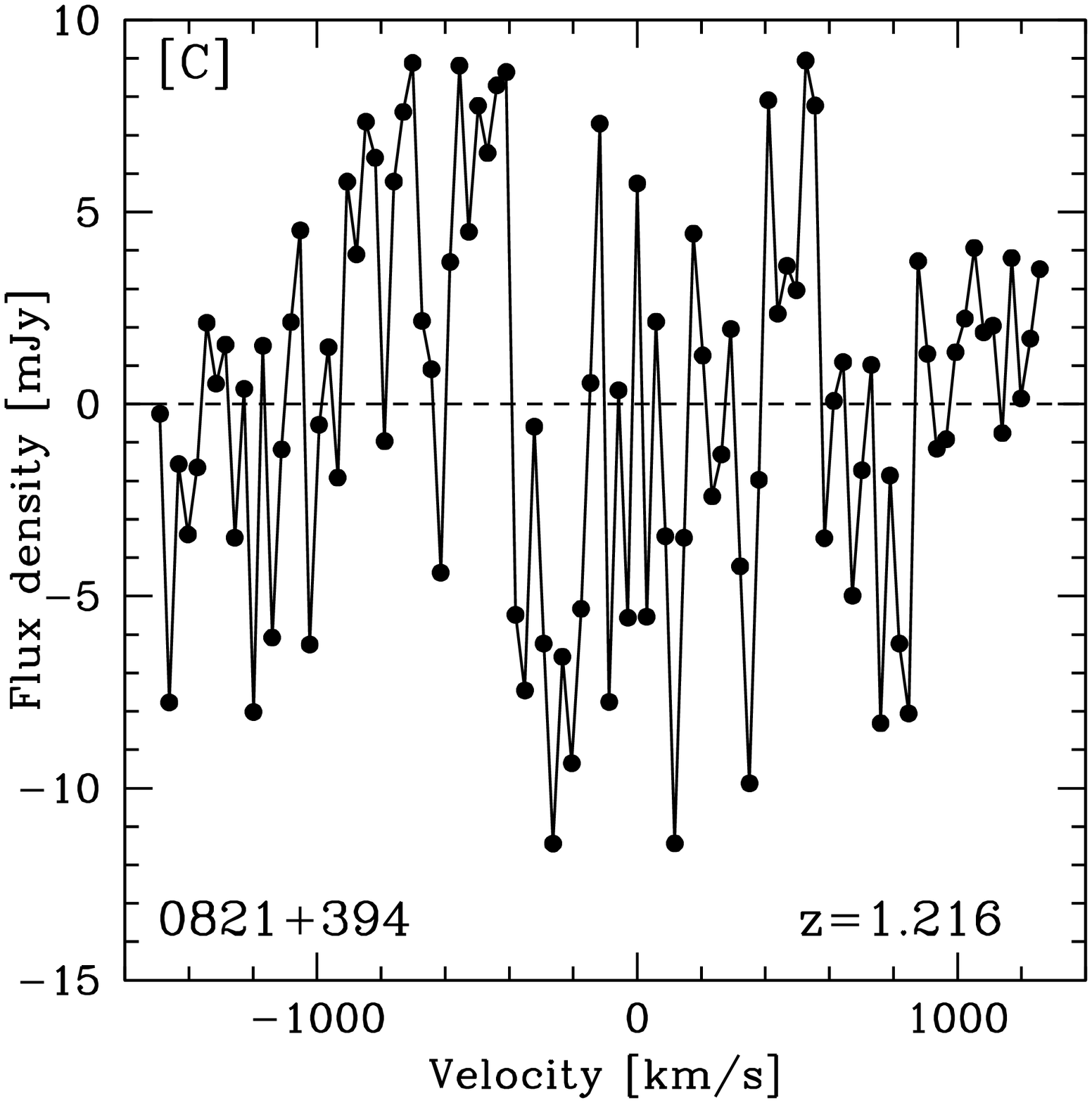} \\
\includegraphics[scale=0.25]{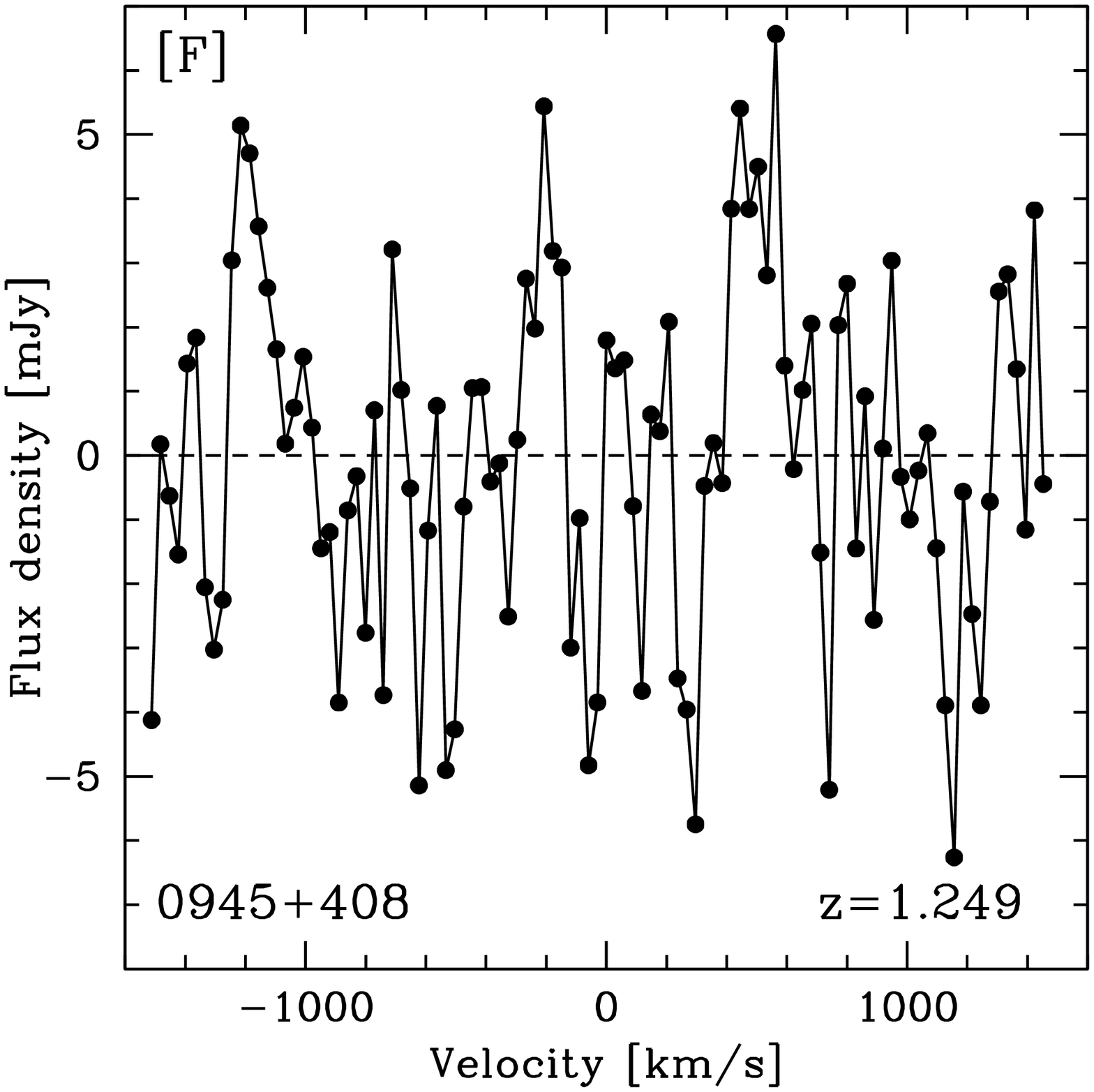} & \includegraphics[scale=0.25]{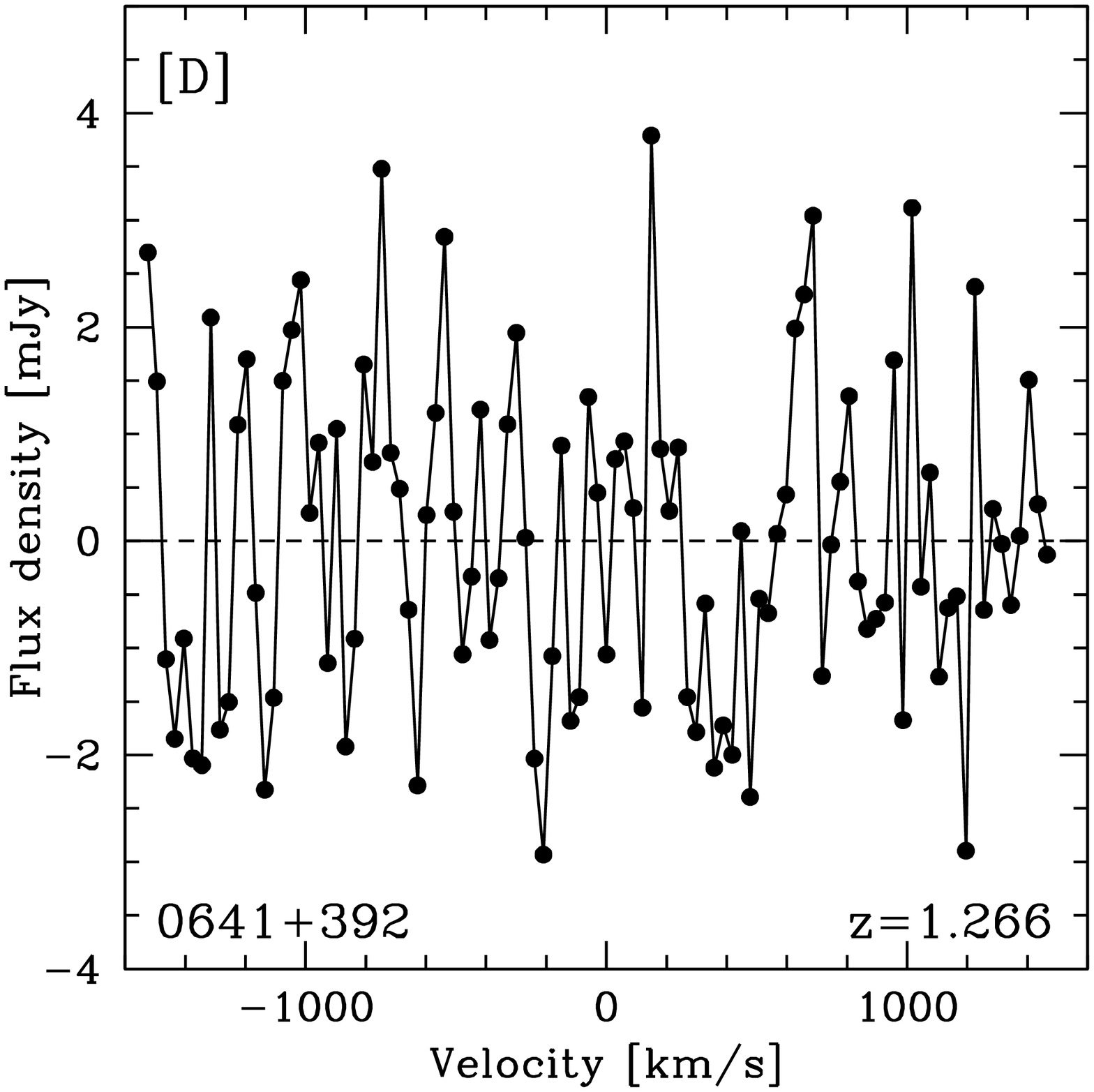} & \includegraphics[scale=0.25]{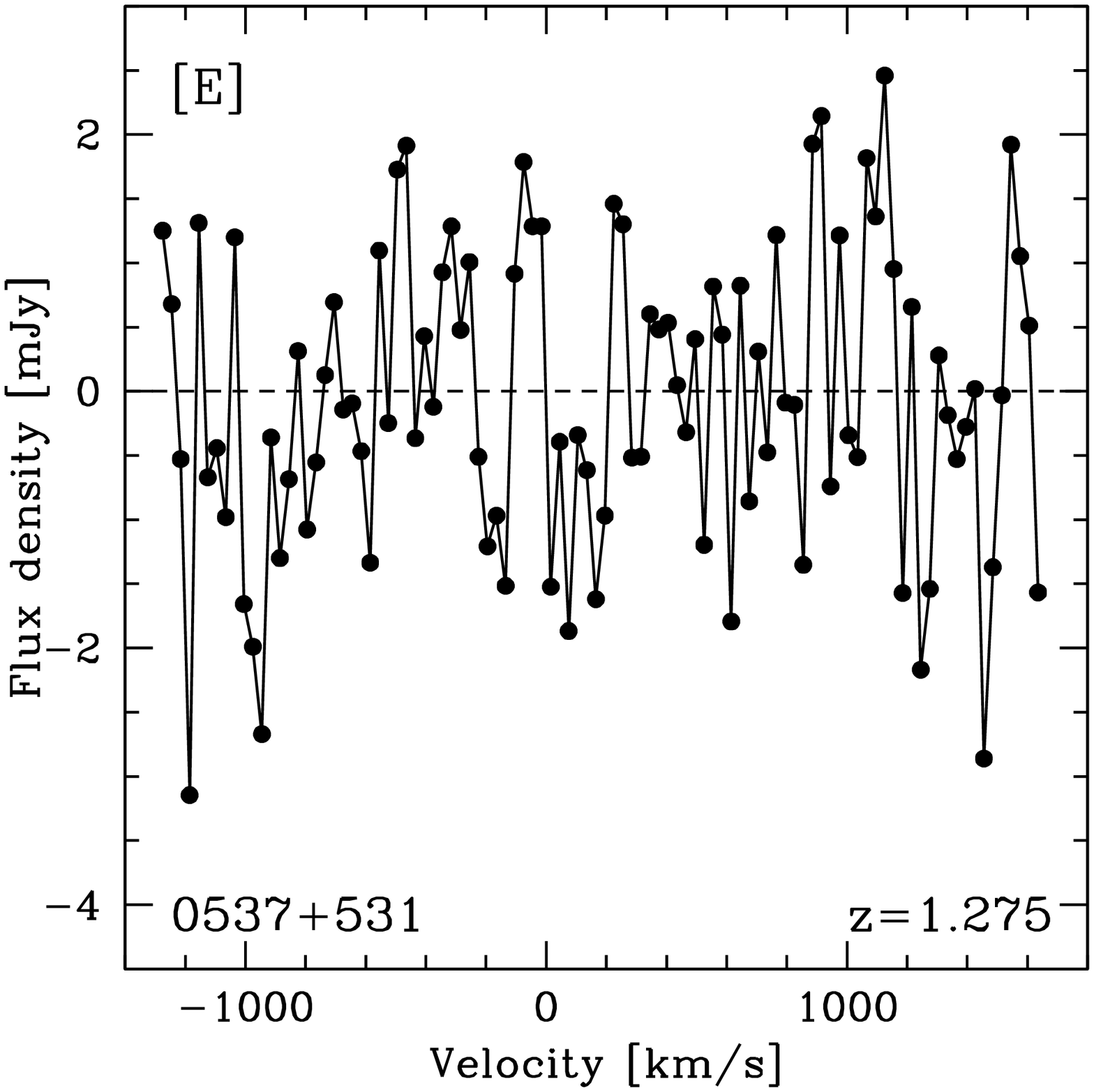} \\
\includegraphics[scale=0.25]{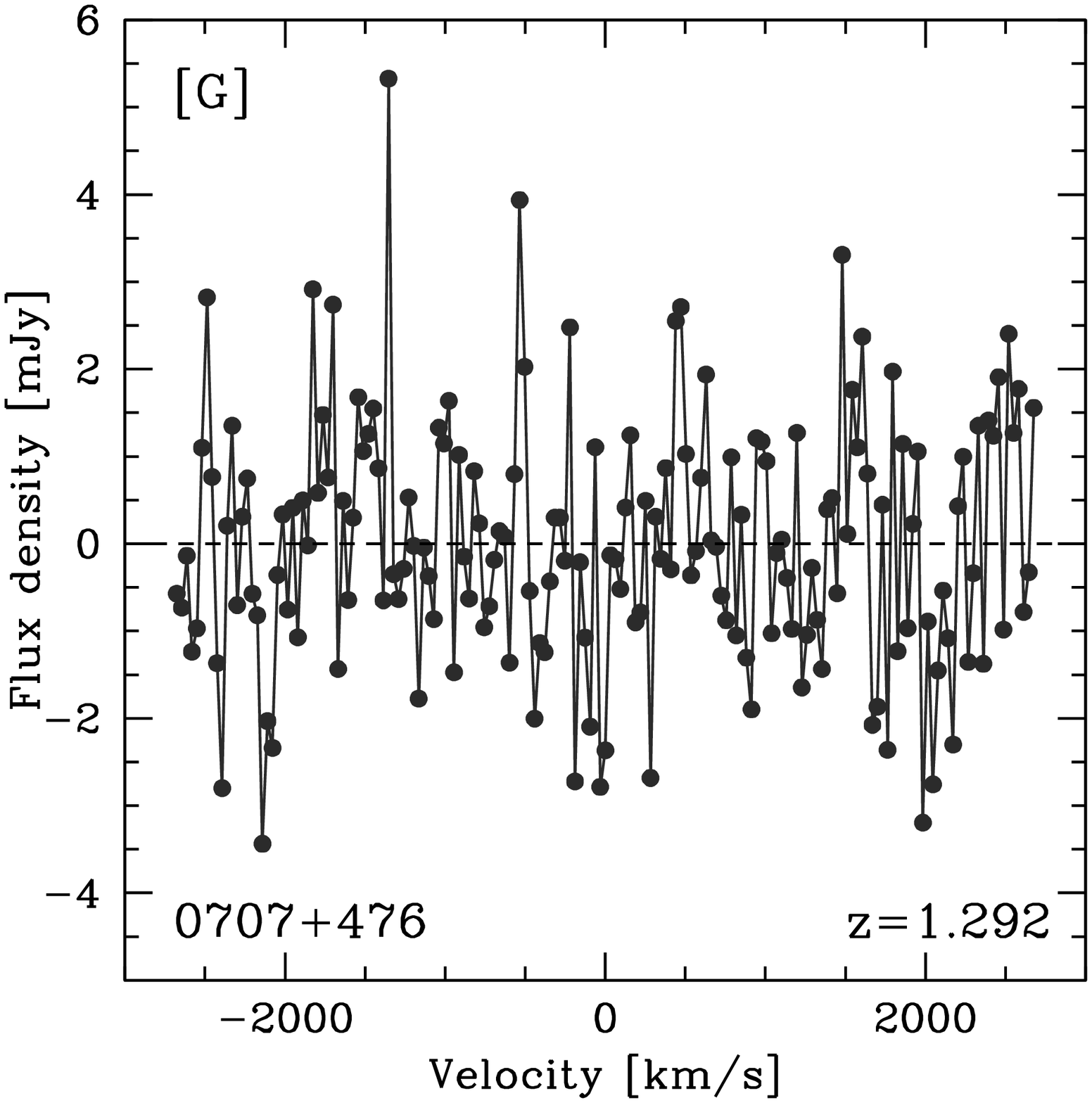} & \includegraphics[scale=0.25]{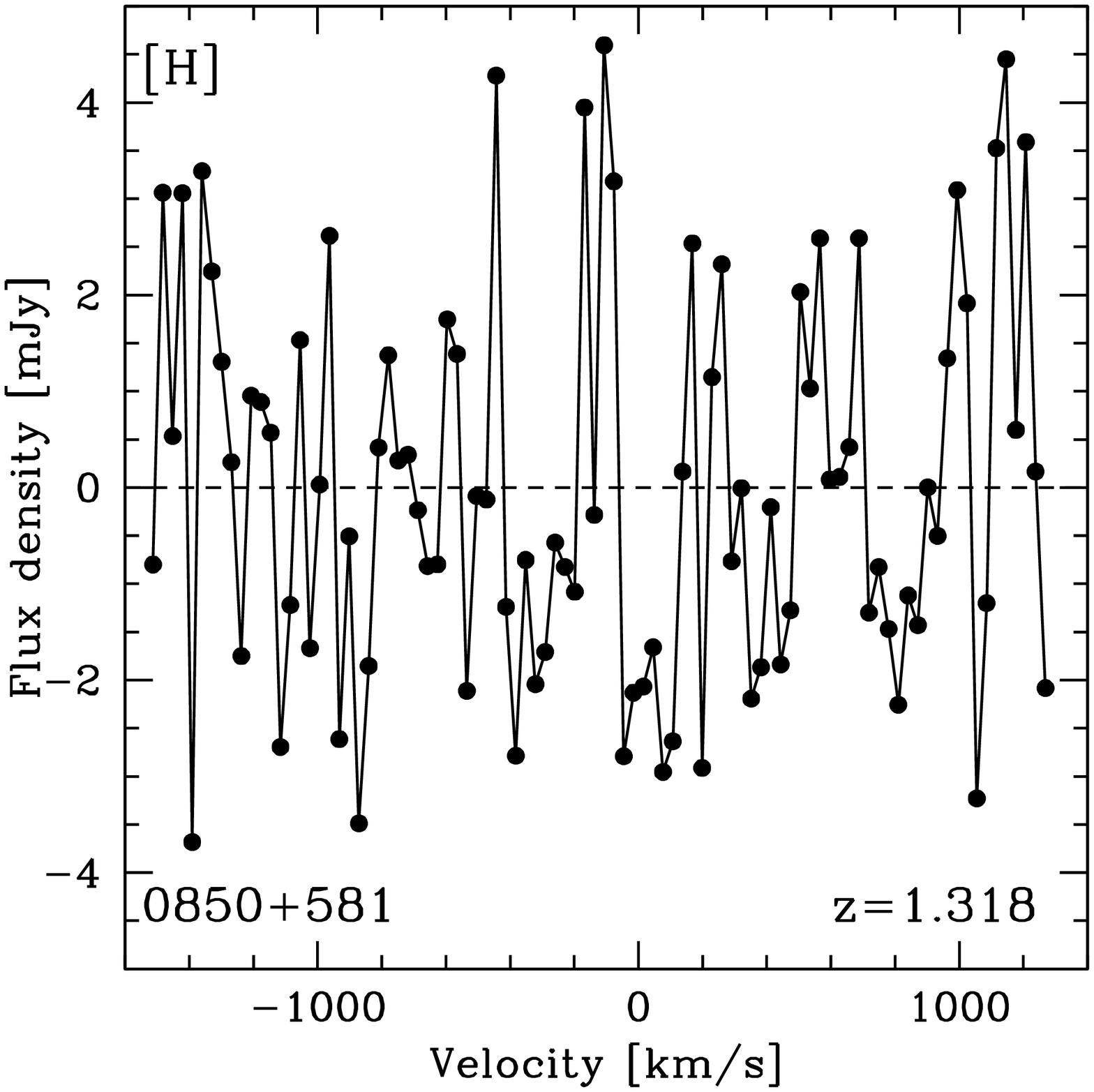} & \includegraphics[scale=0.25]{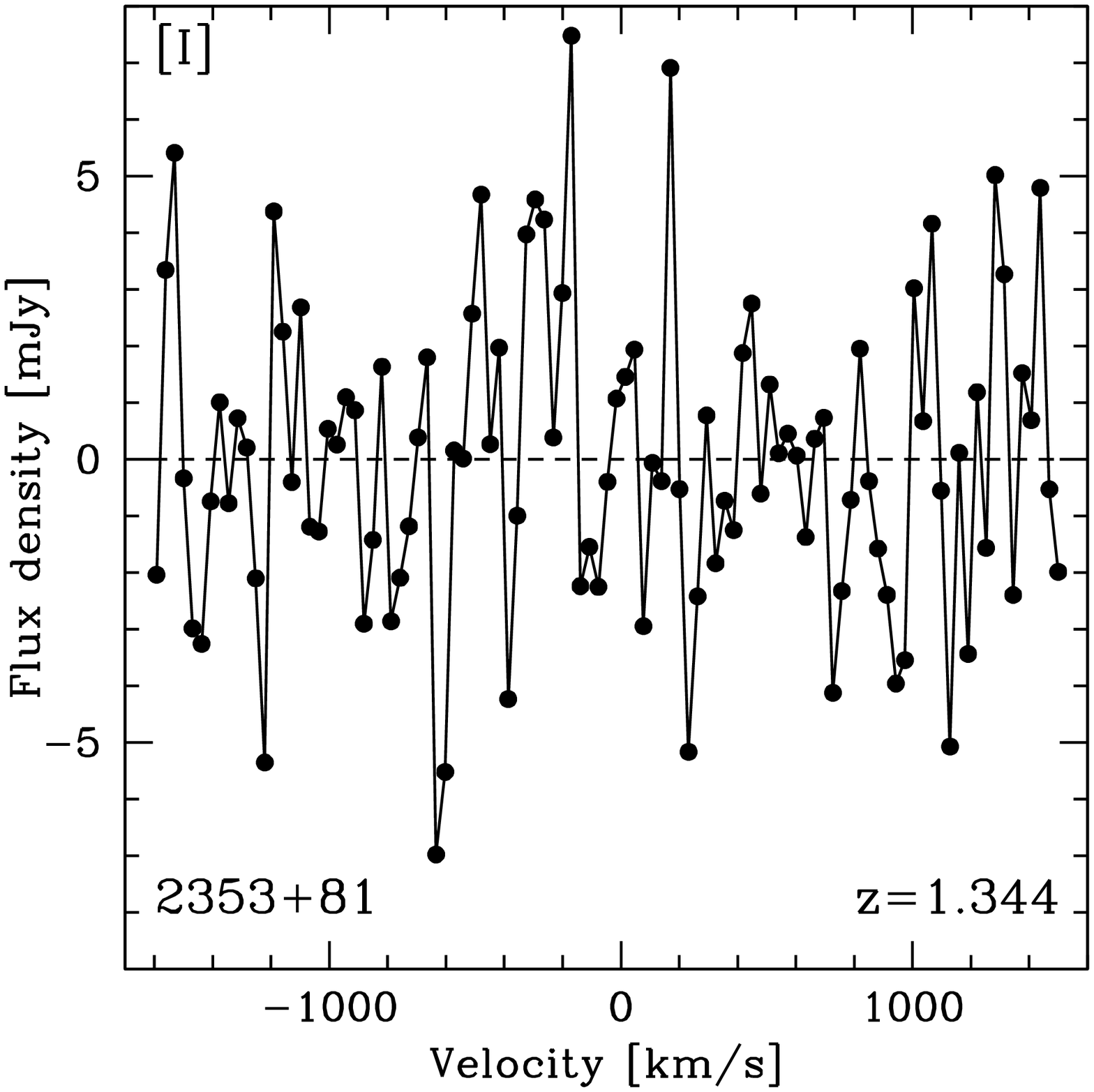} \\
\includegraphics[scale=0.25]{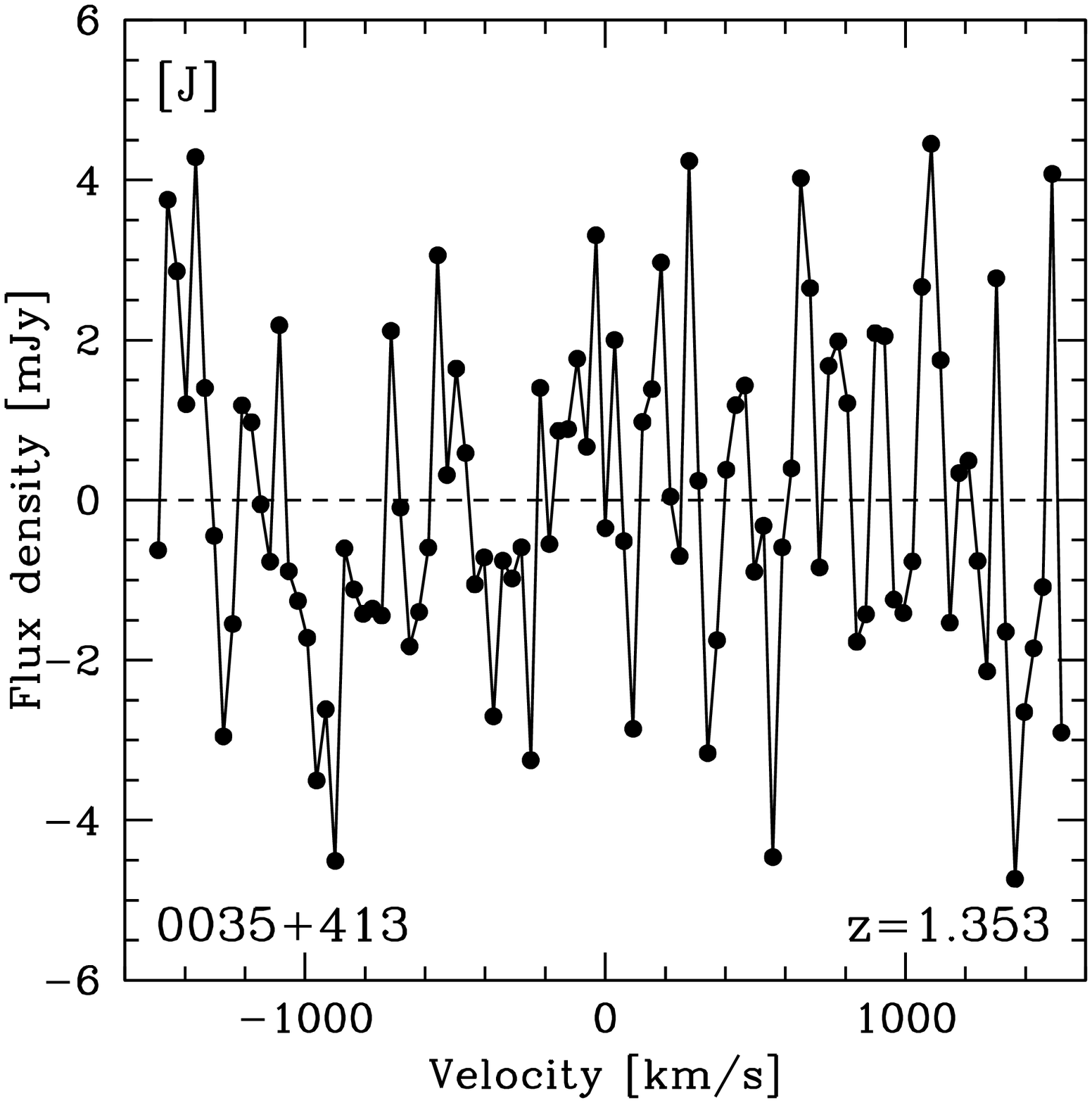} & \includegraphics[scale=0.25]{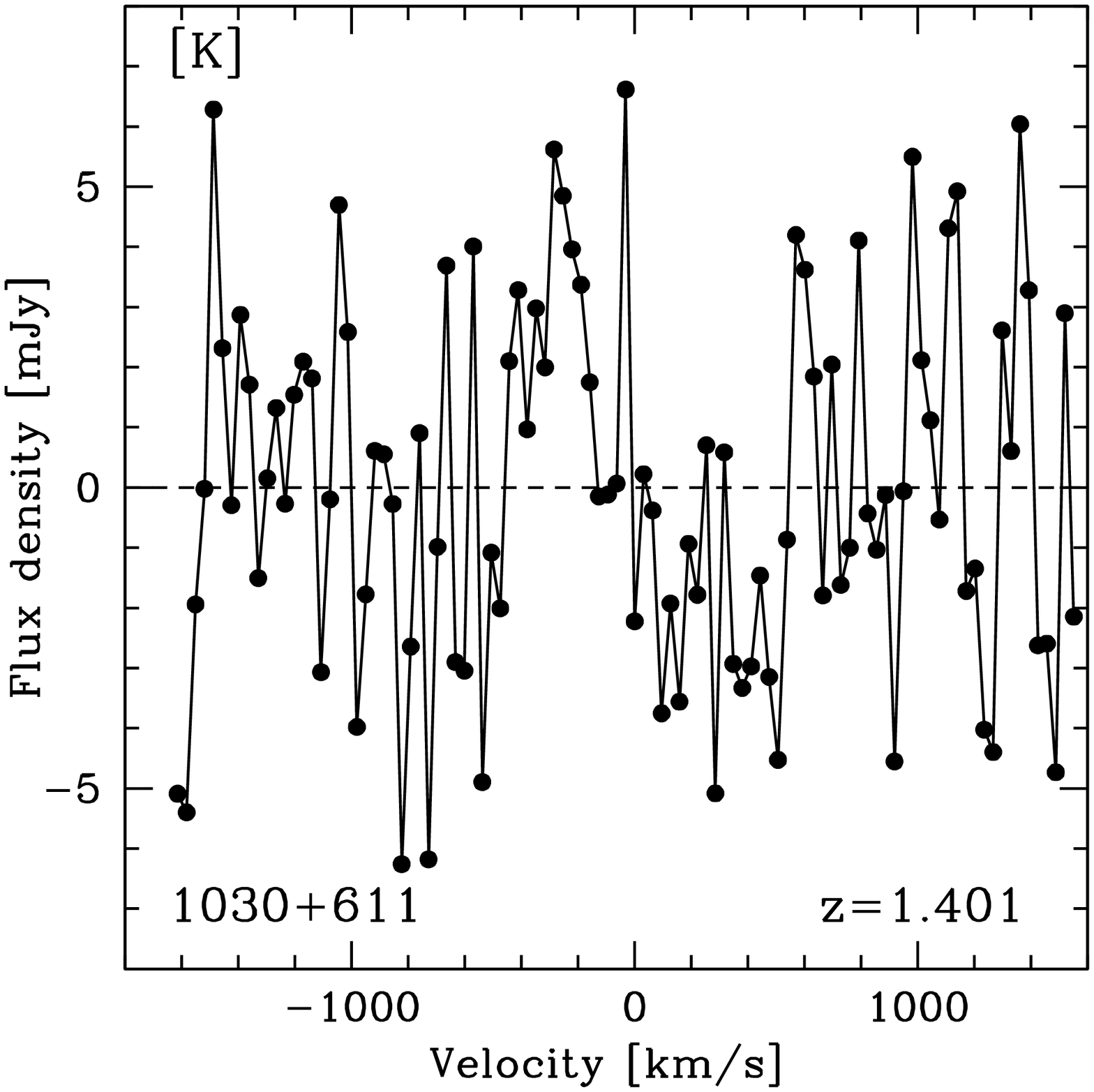} & \includegraphics[scale=0.25]{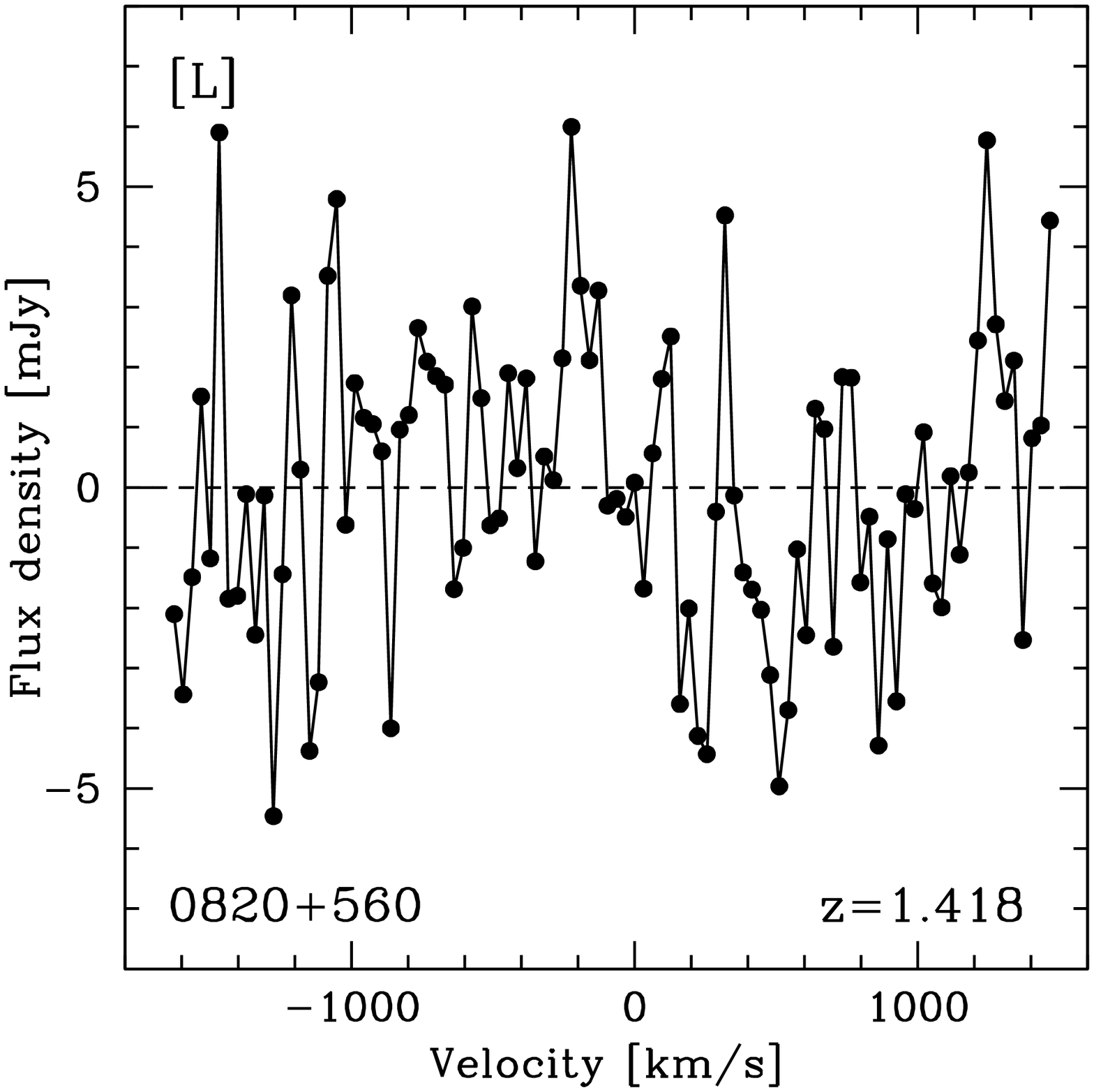} \\
\end{tabular}
\caption{The GMRT \hii\ spectra for the 23 sources with usable data; all spectra have been Hanning-smoothed 
and re-sampled. The shaded channels in the 
spectrum of the source TXS\,2356+390 are corrupted by RFI.
\label{fig:full}}
\end{figure*}

\begin{figure*}
\setcounter{figure}{0}
\centering
\begin{tabular}{ccc}
\includegraphics[scale=0.25]{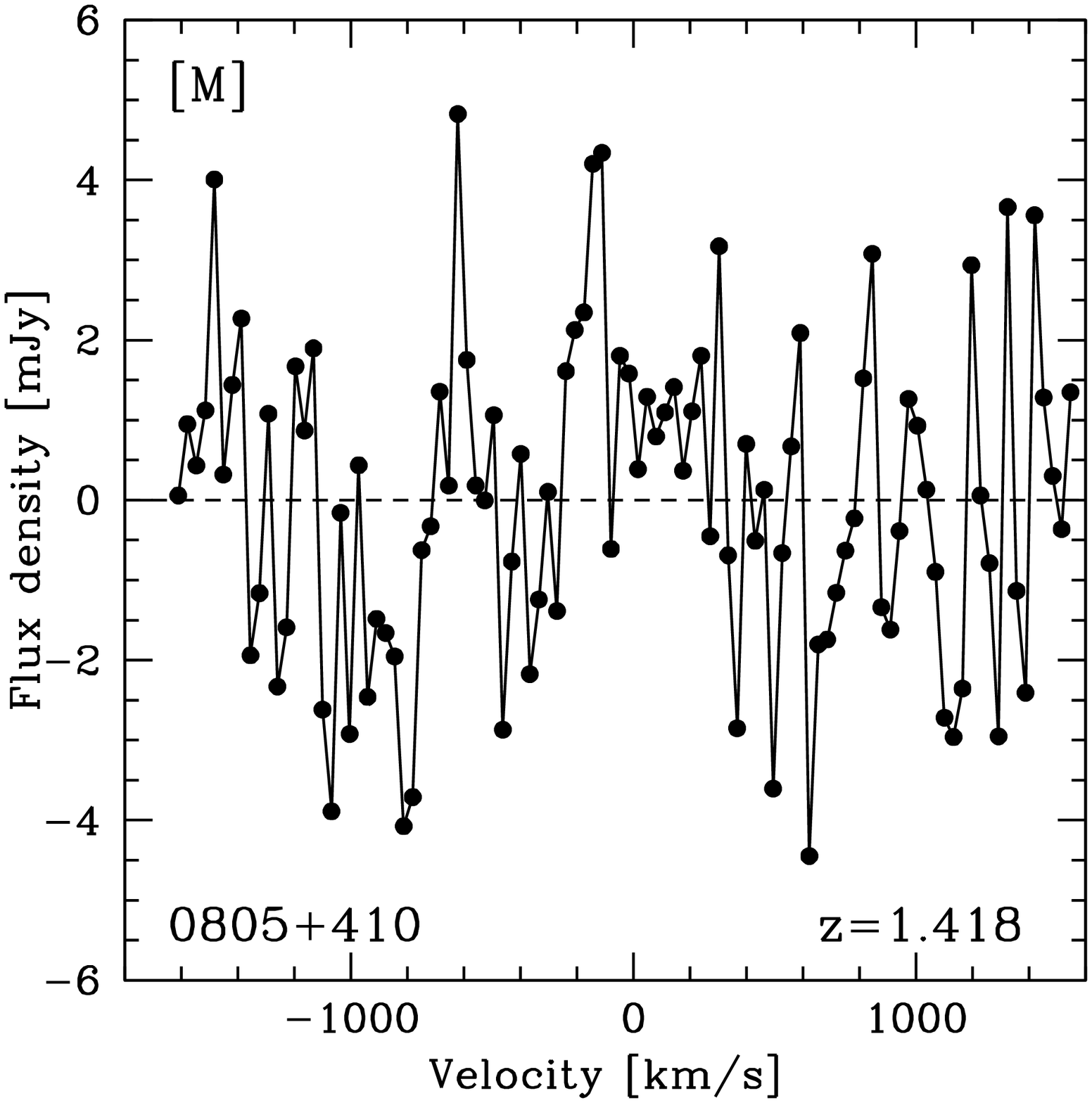} & \includegraphics[scale=0.25]{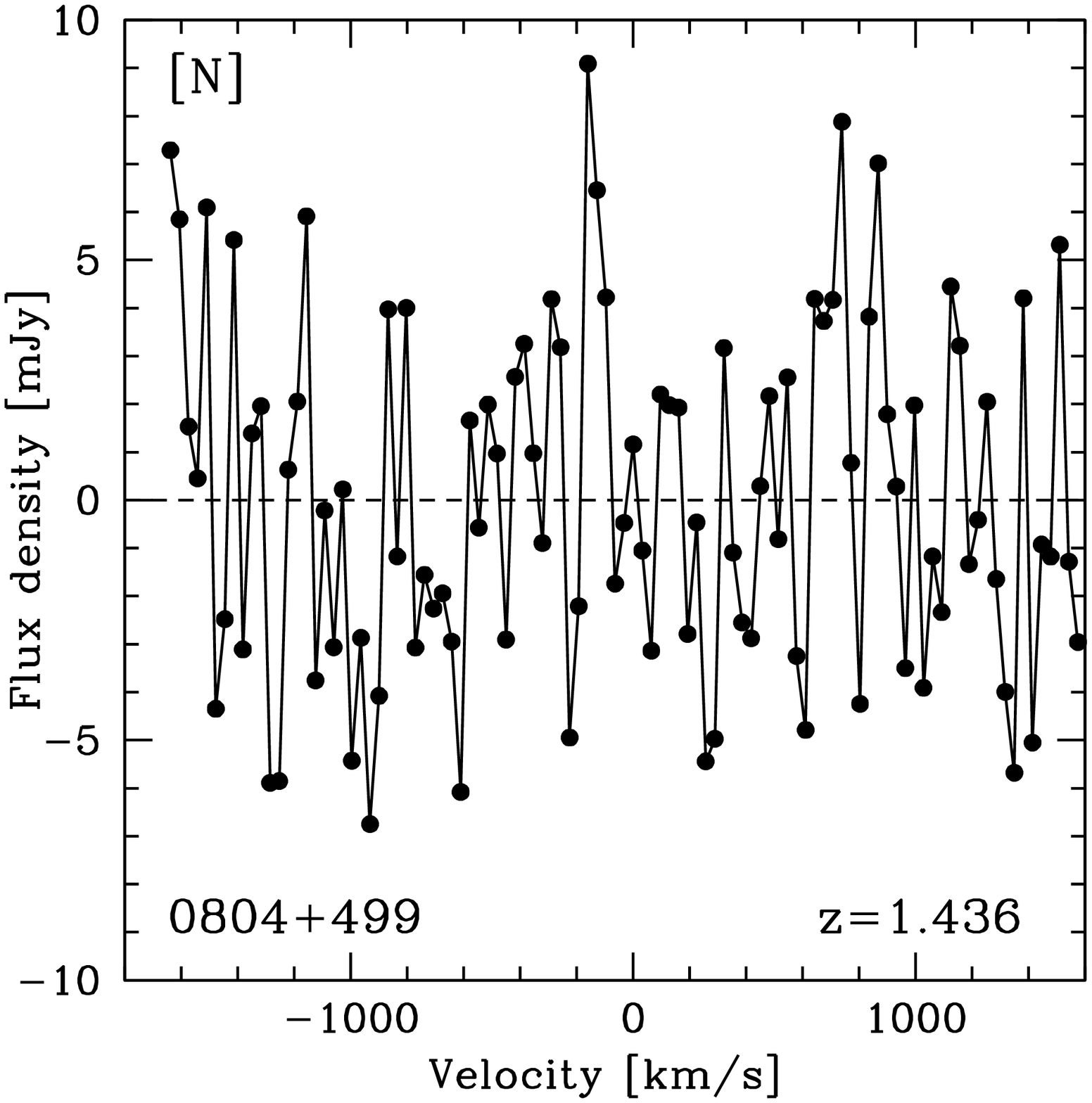} & \includegraphics[scale=0.25]{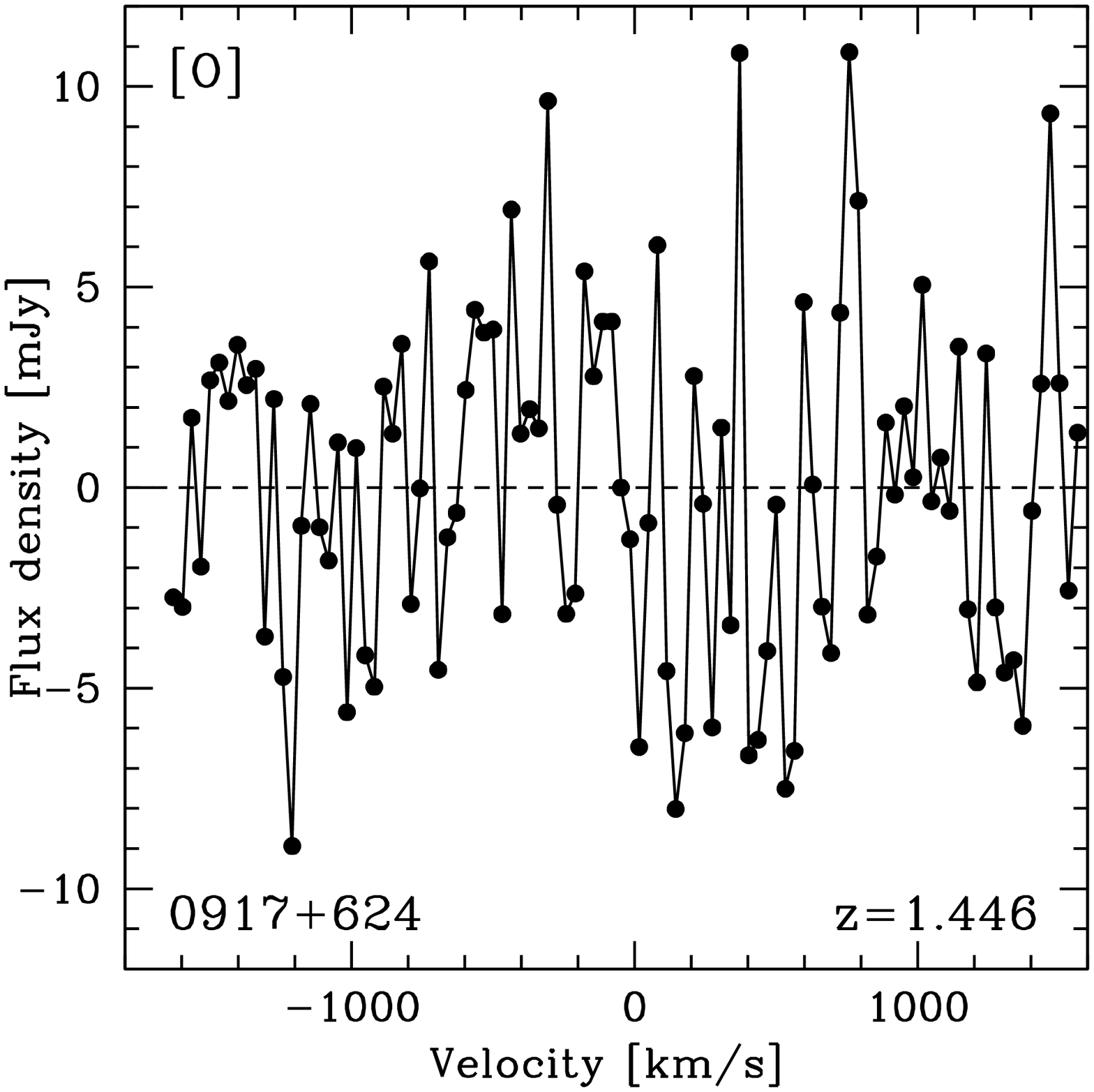} \\
\includegraphics[scale=0.25]{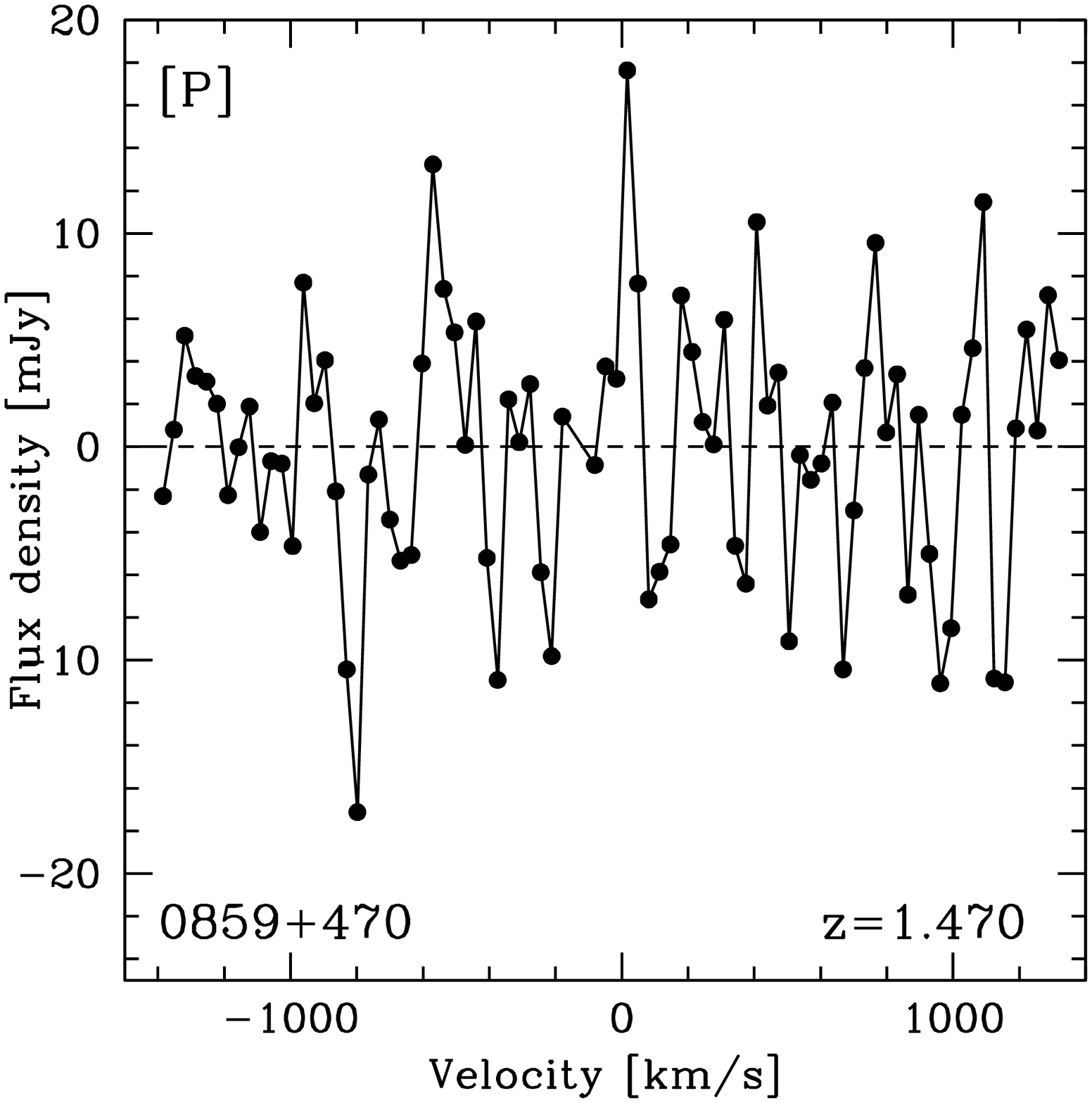} & \includegraphics[scale=0.25]{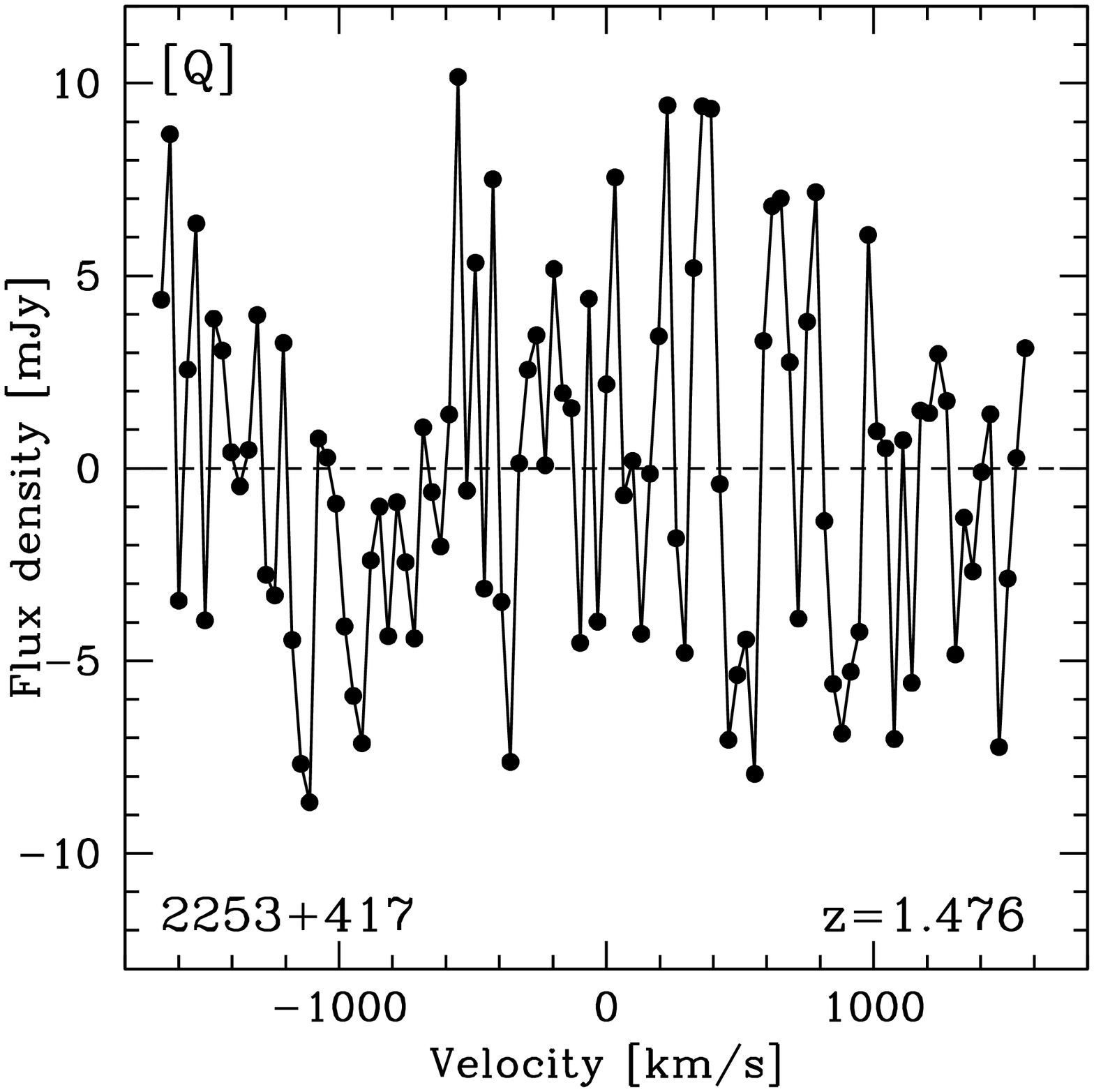} & \includegraphics[scale=0.25]{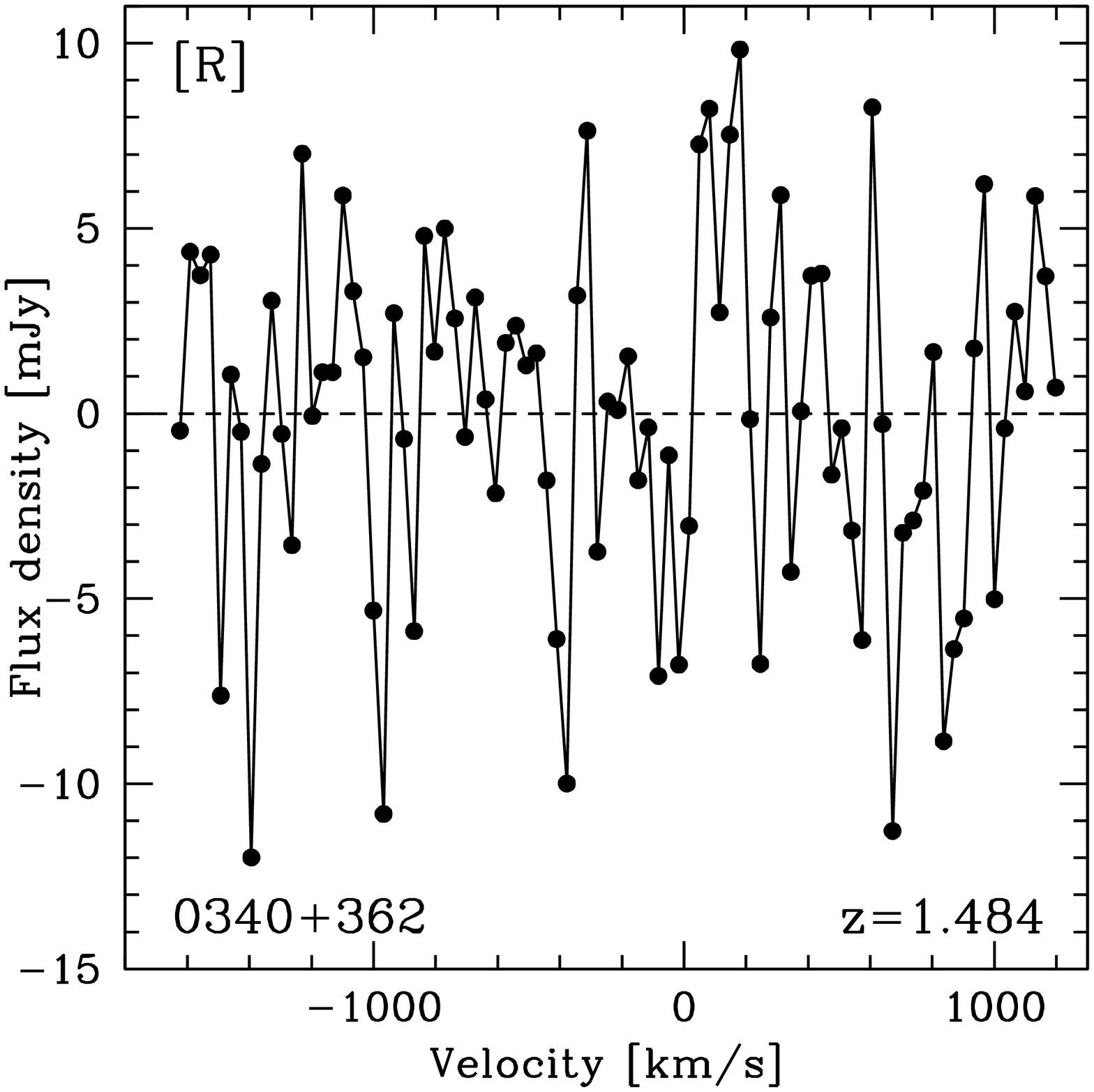} \\
\includegraphics[scale=0.25]{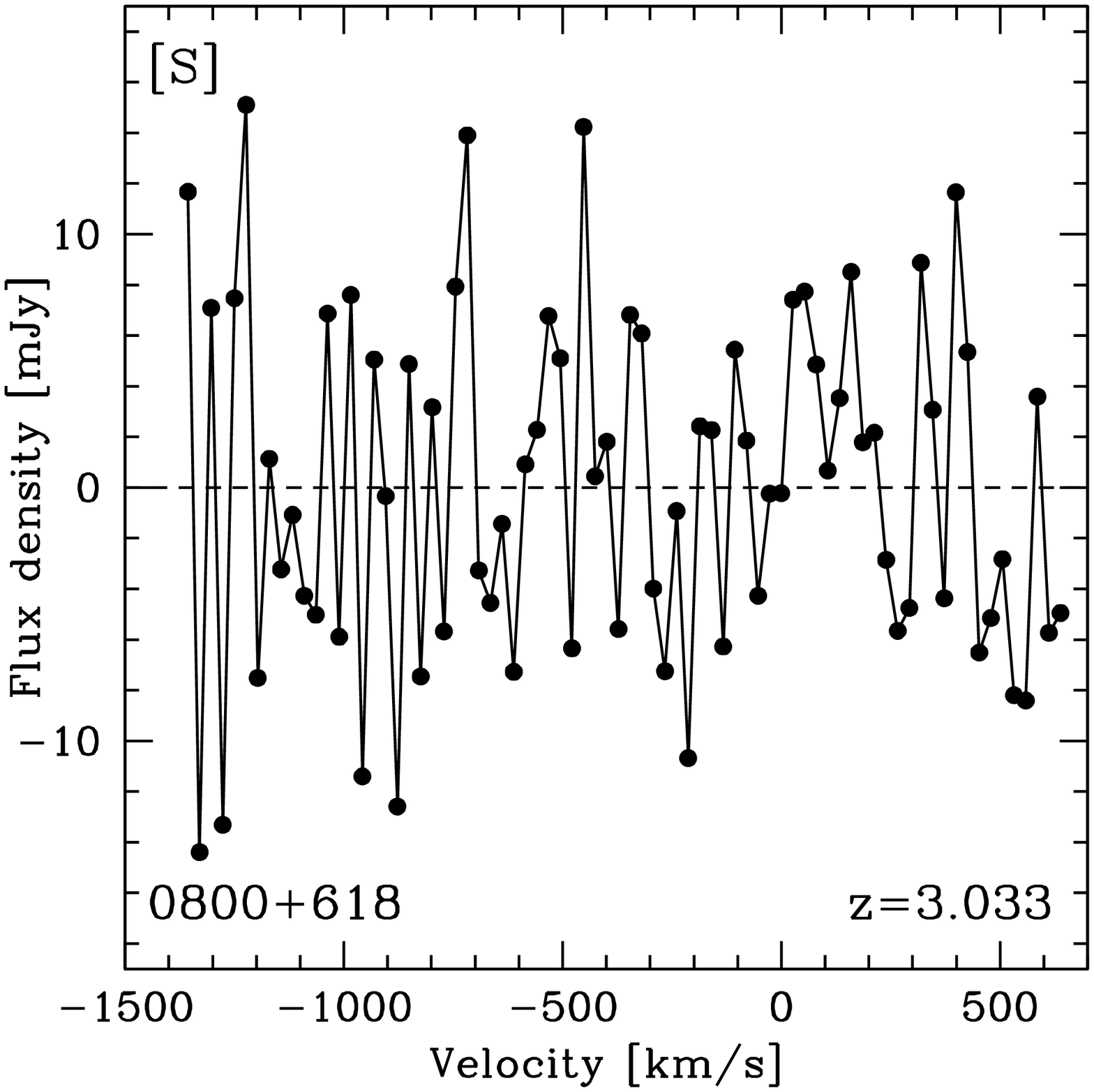} & \includegraphics[scale=0.25]{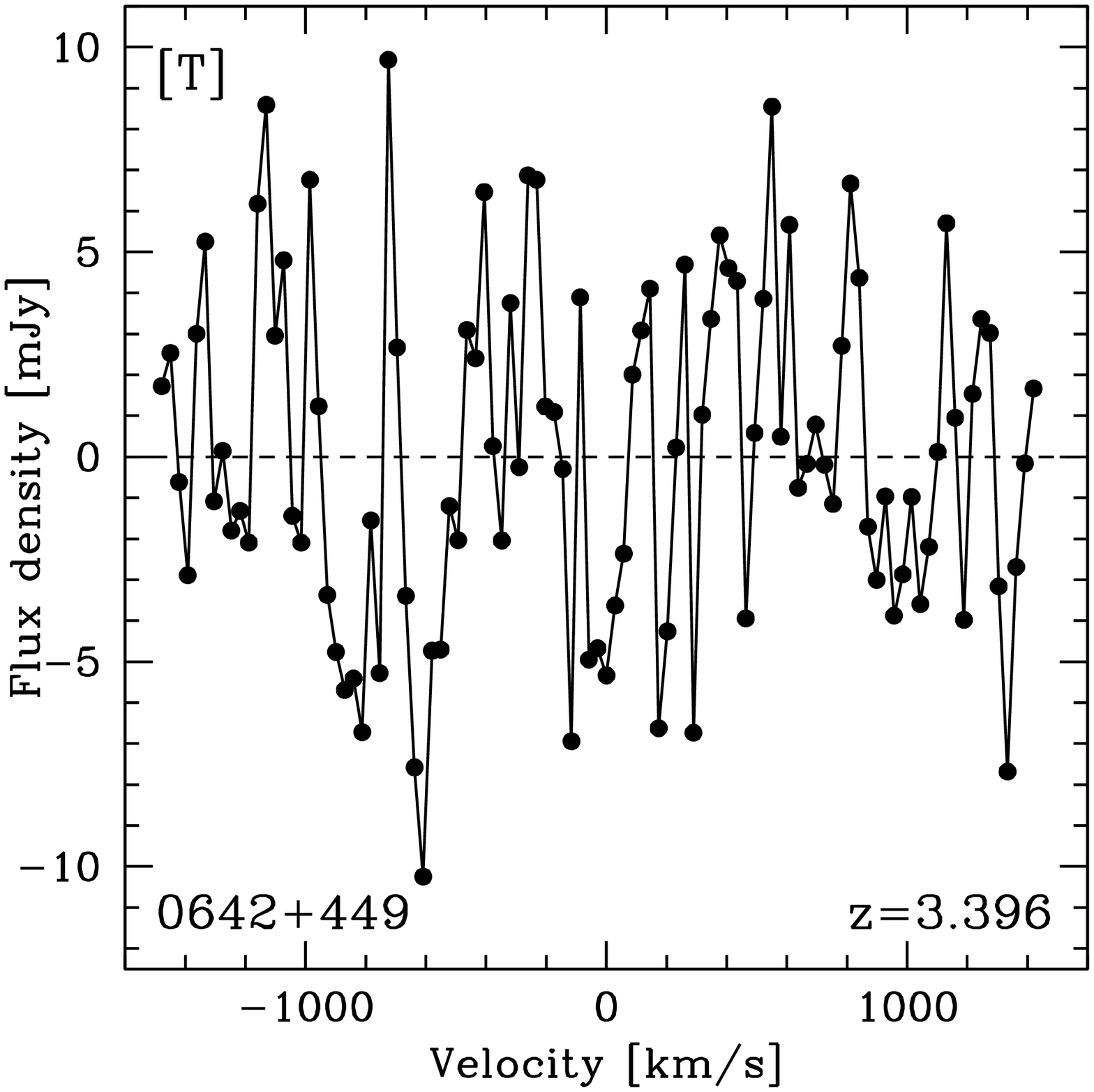} & \includegraphics[scale=0.25]{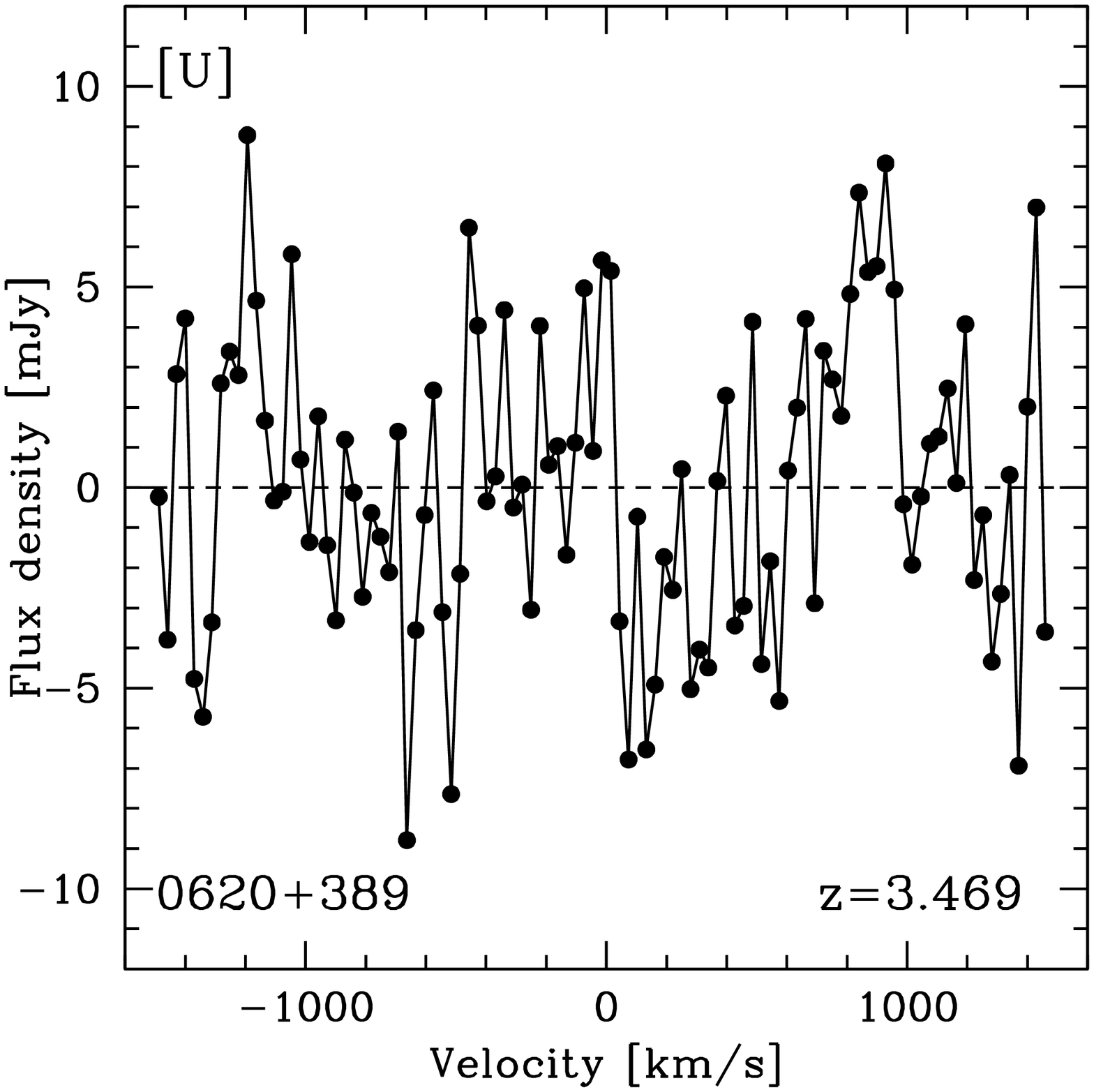} \\
\includegraphics[scale=0.25]{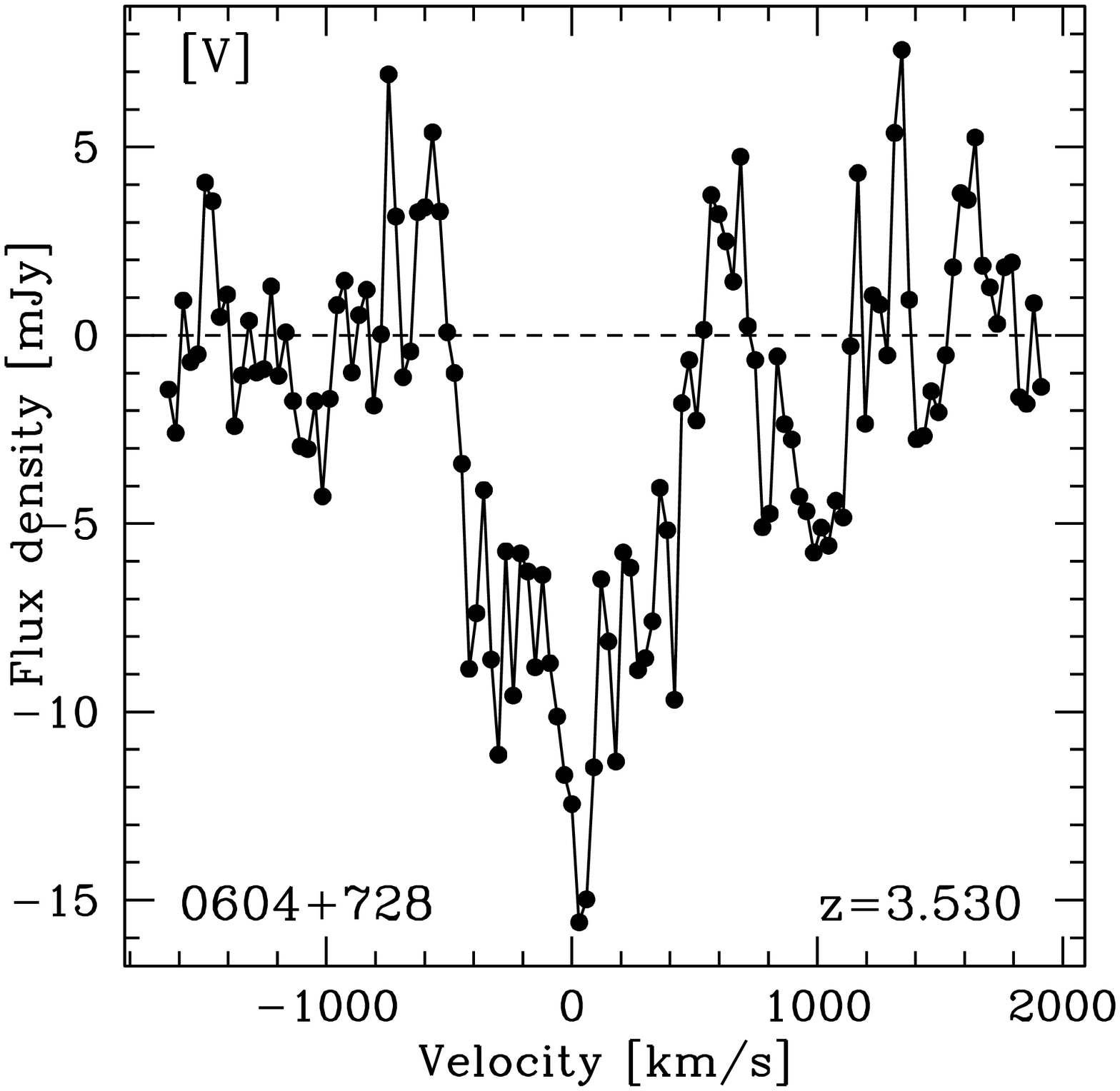} & \includegraphics[scale=0.25]{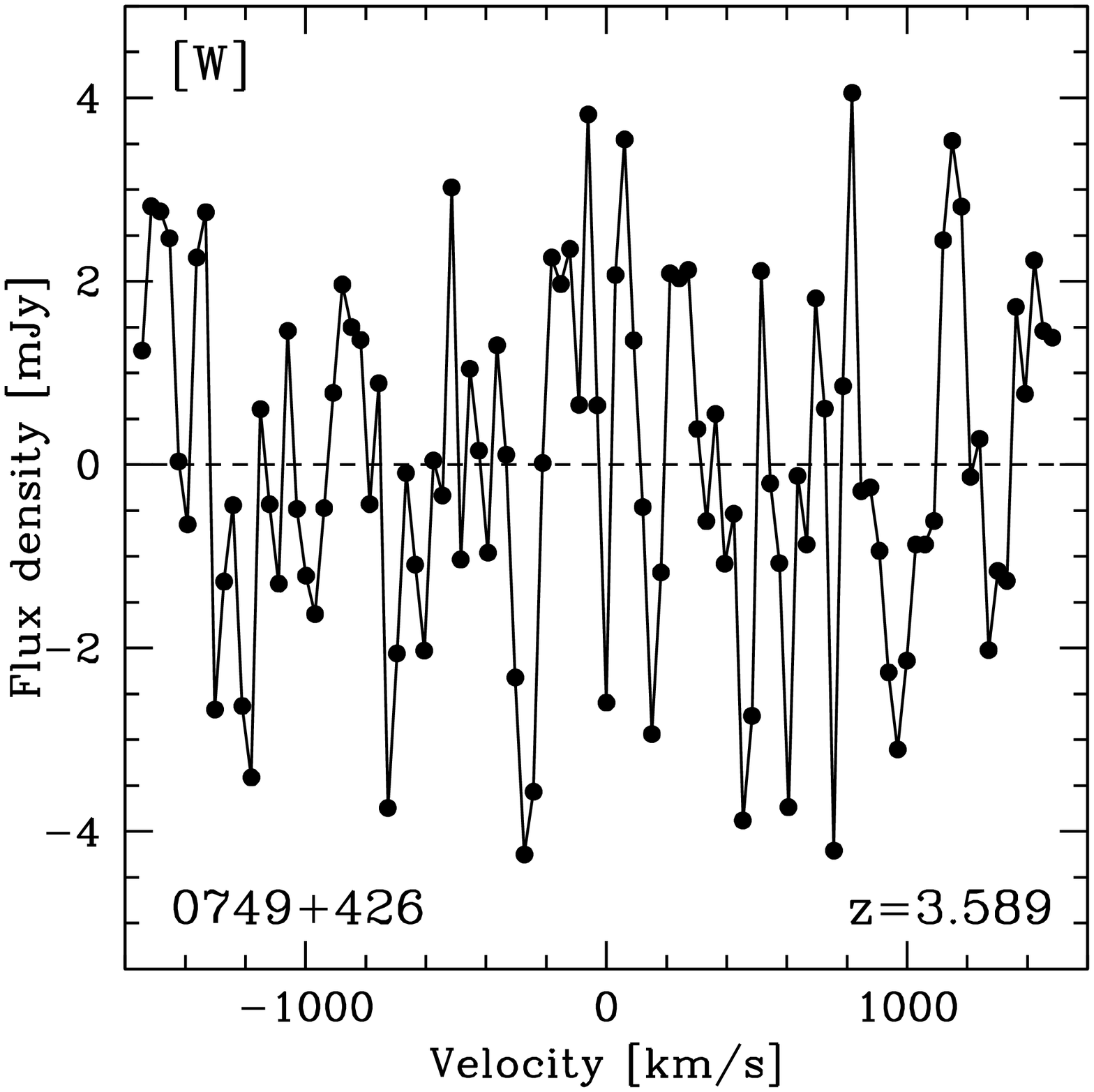} \\ 
\end{tabular}
\caption{(contd.)
\label{fig:full-contd}}
\end{figure*}

\subsection{The Caltech-Jodrell Bank flat-spectrum sample}

A number of studies, mostly targetting AGNs at low redshifts, $z \lesssim 1$, have established that the 
integrated \hii\ optical depth is inversely correlated with the spatial extent of the radio emission 
\citep[e.g.][]{pihlstrom03,gupta06a}. For example, \citet[][]{gupta06a} found that 
the detection rate of \hii\ absorption in their sample of 96 sources was highest for the most compact, 
gigahertz-peaked-spectrum (GPS) sources, and lowest for steep-spectrum sources with extended radio continua. 
We hence used source compactness as the primary criterion in selecting our target AGN sample. Further,
both inverted-spectrum (e.g. GPS) and flat-spectrum AGNs are expected to be relatively compact, as the 
flattening of the radio spectrum at low frequencies is believed to arise from synchrotron 
self-absorption \citep[e.g.][]{odea98}. We hence decided to target flat-spectrum sources as these 
provide a balance between compactness and sufficient low-frequency flux density for deep searches
for redshifted \hii\ absorption.

The Caltech-Jodrell Bank flat-spectrum (CJF) sample \citep[][]{pearson88,polatidis95,henstock95,taylor96} 
was used to select our target AGNs, for the following reasons: (1)~the flat-spectrum criterion implies 
that the sources of the CJF sample are extremely compact, (2)~accurate redshifts are available for most 
of the CJF sample, from follow-up optical spectroscopy \citep[e.g.][]{henstock97}, (3)~the CJF sources are all 
relatively bright at radio frequencies, with 5~GHz flux densities $> 350$ mJy \citep[e.g.][]{taylor96}, 
(4)~VLBI information is available for all CJF sources, at frequencies of $\approx 1-5$~GHz, providing additional 
details of their spatial structure \citep[e.g.][]{polatidis95,taylor96}, and (5)~low-frequency flux density 
estimates are available for all CJF sources from the 325~MHz Westerbork Northern Sky Survey (WENSS) 
\citep[][]{rengelink97} or the 365~MHz Texas survey \citep[][]{douglas96}.

For the pilot GMRT \hii\ absorption survey reported here, we chose to focus on the brightest CJF 
sources, and hence imposed the additional criterion that the WENSS 325~MHz flux density of all targets 
be greater than 500~mJy. This was done to ensure that strong optical depth limits could be achieved 
in relatively short GMRT integration times in this pilot project. A total of 24 AGNs, 6 at $3.0 < z < 3.6$ 
and 18 at $1.1 < z < 1.5$, were included in the pilot sample. We note, in passing, that there are only 
2 AGNs from the CJF sample with searches for associated \hii\ absorption at $z \gtrsim 1$ in the 
literature \citep[e.g.][]{gupta06a,curran13}.


\subsection{The GMRT observations and data analysis}

The GMRT was used to observe the 24 CJF targets in November 2008 (proposal 15NKa01), with the 327~MHz 
and the 610~MHz receivers for sources at $3.0 < z < 3.6$ and $1.1 < z < 1.5$, respectively, and with the 
GMRT hardware correlator as the backend. The observations used bandwidths of 4~MHz (for the 327~MHz band) 
and 8~MHz (for the 610~MHz band), centred at the redshifted \hii\ line frequency of each target, and 
subdivided into 256~channels, using the high-resolution mode 
of the hardware correlator. This provided a total velocity coverage of $\approx 3200 - 3900$~\kms\ 
and a velocity resolution of $\approx 12 - 15 $~\kms. The typical on-source times were $\approx 1.0$~hr 
and $\approx 3$~hrs apiece, for sources observed with the 610~MHz and the 327~MHz bands, respectively. 
Observations of the standard flux calibrators 3C48, 3C147 or 3C286 were used to calibrate the 
GMRT's flux density scale, while (for most targets) nearby compact sources were used as secondary gain 
calibrators.

The initial GMRT searches yielded two absorption features, at $z = 1.198$ towards TXS\,2356+390, 
and $z=3.530$ towards TXS\,0604+728. The first of these was a weak feature (with $\approx 5\sigma$ 
significance, after integrating over the profile), while the second had high ($\gtrsim 20\sigma$) 
significance. In addition, data on two targets (TXS\,0707+476 and TXS\,0014+813) were affected by RFI. We hence used 
the GMRT with its new software backend and 512 channels to re-observe all four systems in 2013 and 2014, 
using bandwidths of 4.17~MHz for TXS\,0604+728 and 16.7~MHz for TXS\,0014+813 (both observed with the 327~MHz band) 
and of 33.33~MHz for TXS\,2356+390 and TXS\,0707+476 (both observed with the 610~MHz band). The on-source 
times were again $\approx 1.0$~hr (610~MHz band) and $\approx 3$~hrs (327~MHz band) apiece. 

All GMRT data were analysed in ``classic'' {\sc aips}, using standard procedures. For each 
target source, the data were first carefully inspected and edited to remove non-working 
antennas, bad correlator baselines, and time-specific problems, the latter usually arising due
to intermittent radio frequency interference (RFI). Next, after calibration of the antenna-dependent 
gains and bandpass shapes, an iterative self-calibration procedure was followed for each target, 
consisting of (typically) 3-4 rounds of phase-only self-calibration and imaging, followed by 1-2 
rounds of amplitude-and-phase self-calibration and imaging, with additional data editing to remove 
corrupted data identified during the above process. This was carried out until the procedure converged 
to yield an image of the target that did not improve on further self-calibration. Since all our targets 
are unresolved at the GMRT angular resolution ($\approx 7''$ at 610~MHz and $\approx 12''$ at 327~MHz),
the task {\sc jmfit} was then used to measure the flux density of each target via a single-Gaussian fit 
to a small region around the target source in the image plane. The continuum image was then subtracted 
from the calibrated U-V visibilities using the tasks {\sc uvsub} and {\sc uvlin}. Additional data 
editing was carried out at this stage, involving a detailed inspection of the residual dynamic 
spectra on all baselines. The residual U-V visibilities were then imaged to produce a ``dirty'' 
spectral cube, and a spectrum was obtained at the location of the target source via a cut through 
the cube. Finally, in some cases, a second-order polynomial was fit to each spectrum and subtracted 
out, to compensate for residual bandpass effects.

\begin{figure}
\caption{The GMRT \hii\ spectrum for the sole (tentative) detection of \hii\ absorption 
in our sample, at $z = 3.530$ towards TXS\,0604+728. Note that the spectrum is at the original 
velocity resolution, i.e. it has not been Hanning-smoothed and re-sampled.
\label{fig:0604}}
\centering
\includegraphics[scale=0.4]{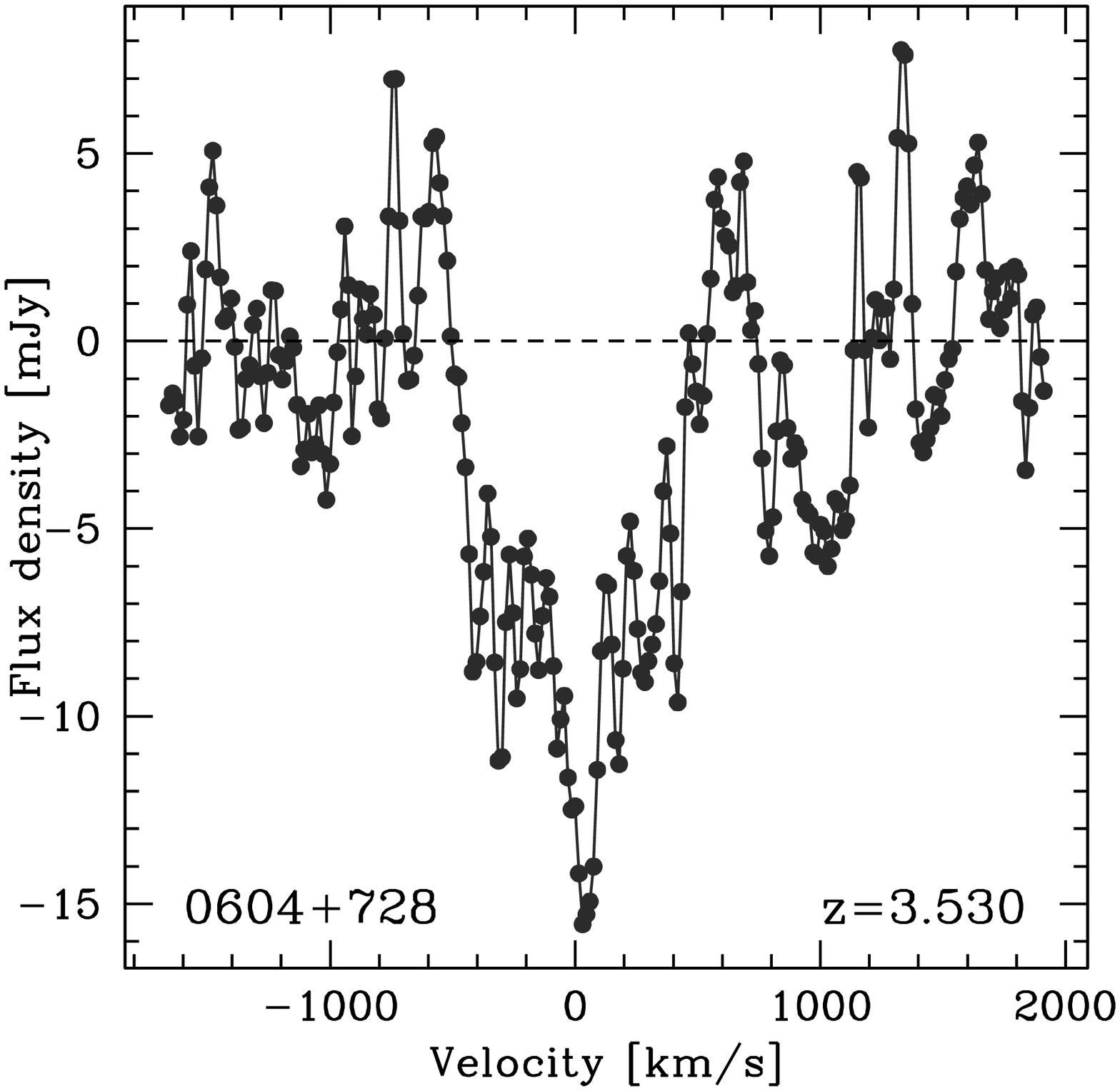}
\end{figure}

\subsection{Results}
\label{sec:results}

The final GMRT \hii\ absorption spectra towards 23 of the 24 sources of the sample are shown 
in Fig.~\ref{fig:full}. Except for the tentative detection of \hii\ absorption towards 
TXS\,0604+728, all spectra have been Hanning-smoothed and re-sampled.  In the case of one source, 
TXS\,0014+813, the GMRT data from multiple observing 
runs were affected by RFI and it was not possible to obtain a reliable RFI-free spectrum; this
source will hence be excluded from the later discussion. For 21 sources, the GMRT spectra were 
consistent with noise, with no evidence for statistically significant absorption features.
For two sources, TXS\,2356+390 at $z = 1.198$ and TXS\,0604+728 at $z=3.530$, weak absorption features 
(with $\geq 5\sigma$ significance) were seen in the initial GMRT spectra. In the case of TXS\,2356+390, 
the follow-up GMRT observations (with somewhat worse sensitivity) ruled out the presence of \hii\ absorption 
at the level detected in the first observing run, indicating that the original feature is likely to 
have arisen due to low-level RFI. In the case of TXS\,0604+728, at $z=3.530$, the data from multiple 
GMRT observing runs following the original 
tentative detection were affected by strong RFI, and it was hence not possible to confirm or rule 
out the reality of the absorption feature. We note that the absorption feature towards TXS\,0604+728 was 
detected via independent analysis procedures carried out by two of the authors, with independent data editing,
and was also seen in the two independent polarizations, with consistent strengths. However, we can still not 
formally rule out the possibility that it arises from low-level RFI and hence will refer to it in the 
later discussion as a tentative detection. The final \hii\ spectrum towards TXS\,0604+728 is also shown 
separately in Fig.~\ref{fig:0604} and the source is discussed in more detail below.

Table~\ref{table:data} summarizes the observational details and the results from the GMRT observations,
with the sources ordered in increasing redshift. 
The columns of this table are (1)~the source name, (2)~the source redshift, 
(3)~the observing frequency, $\nu_{\rm 21\,cm}$, in MHz, 
(4)~the observing bandwidth, BW, in MHz, (5)~the velocity resolution $\Delta v$ of the final spectra,  in
\kms\ (in all cases but one, TXS\,0707+476,  after Hanning-smoothing and re-sampling), 
(6)~the source flux density, $S_\nu$, measured using {\sc jmfit}, in mJy, 
(7)~the RMS noise $\Delta S$ on the final spectrum at the velocity resolution listed in column~(5), 
in mJy, (8)~the integrated \hii\ optical depth $\tau d{\rm V}$ in \kms, or, for non-detections, the $3\sigma$ 
upper limit on $\int \tau d{\rm V}$, 
assuming a line full-width-at-half maximum (FWHM) of 100~\kms, (9)~the ultraviolet (UV)
luminosity at a rest-frame wavelength of $1216$~\AA\ of the AGN, (10)~the rest-frame 1.4~GHz radio luminosity
of the AGN, and (11)~the low-frequency spectral index of the AGN, $\alpha_{\rm 21\,cm}$, around the 
redshifted \hii\ line frequency. Note that the details provided for TXS\,0707+476 are for the wide-band
data set, with a bandwidth of 33.33~MHz and a velocity resolution of 31.5~\kms\ (without Hanning-smoothing
and re-sampling).

For the non-detections, the limits on the integrated \hii\ optical depths were computed for an assumed line 
FWHM of 100~\kms, after smoothing the spectrum to a similar velocity resolution. 
Next, the luminosity at a rest-frame wavelength 
of $1216$~\AA\ was estimated for each AGN following the prescription of \citet{curran08}. For 
each AGN, we first determined the flux density $F_{\rm UV}$ at the wavelength $1216 \times (1 + z)$~\AA\ 
by interpolating between its measured flux densities in different optical and UV wavebands from the 
literature, fitting a power-law spectral shape. The luminosity at the rest-frame wavelength of 
$1216$~\AA\ was then inferred from the expression $L_{\rm UV} = 4\pi D_{\rm AGN}^{2} F_{\rm UV} / (1 + z)$.
For two systems, TXS\,0600+442 and TXS\,2253+417, the luminosity is known only at a single optical 
waveband, quite distant from the redshifted $1216\,\AA$ wavelength; these sources hence do not have 
a listed rest-frame $1216\,\AA$ UV luminosity in the table. Finally, the radio spectral indices
of the AGNs were computed from the flux densities at the redshifted \hii\ line frequency and the closest
frequency with a flux density estimate \citep[usually 1.4~GHz, from the Faint Images of the Radio Sky at 
Twenty-cm survey or the NRAO VLA Sky Survey (NVSS);][]{becker95,condon98}

\begin{figure*}
\caption{[A]~Left panel: The integrated \hii\ optical depths of the 52 CJF sources 
of our full sample plotted versus redshift. The 23 sources whose \hii\ absorption spectra are 
presented in this paper are shown as squares, while the 29 literature sources are shown as 
triangles. Filled symbols indicate detections of \hii\ absorption, while open symbols indicate
upper limits on the \hii\ optical depth. The dashed vertical line indicates the median redshift 
of the sample, $z_{\rm med}= 0.76$. [B]~Right panel: The detection rate of 
\hii\ absorption for the sub-samples with $z > z_{\rm med}$ and $z < z_{\rm med}$.
\label{fig:tau}}
\begin{tabular}{cc}

\includegraphics[scale=0.45]{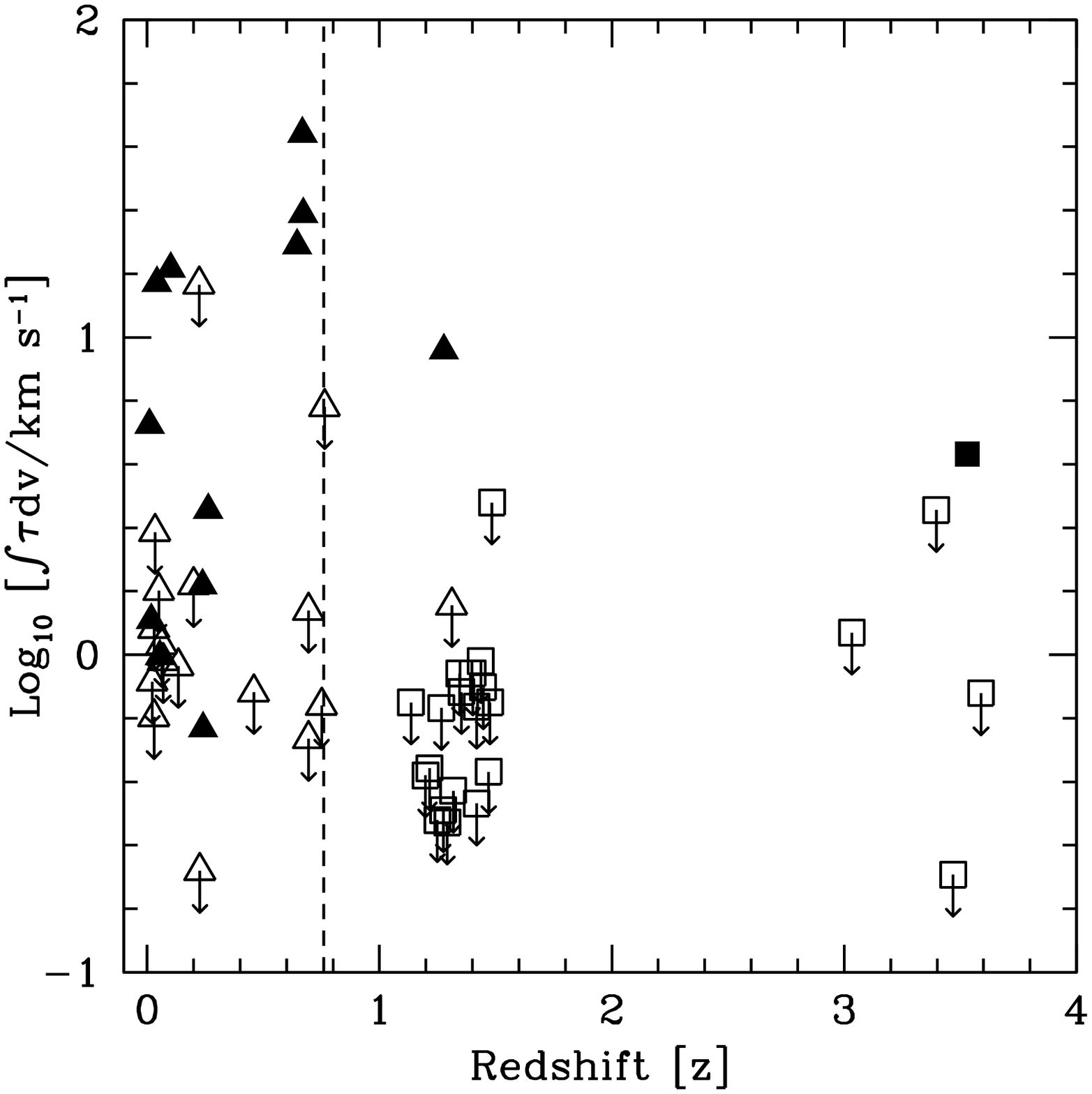}  &  \includegraphics[scale=0.45]{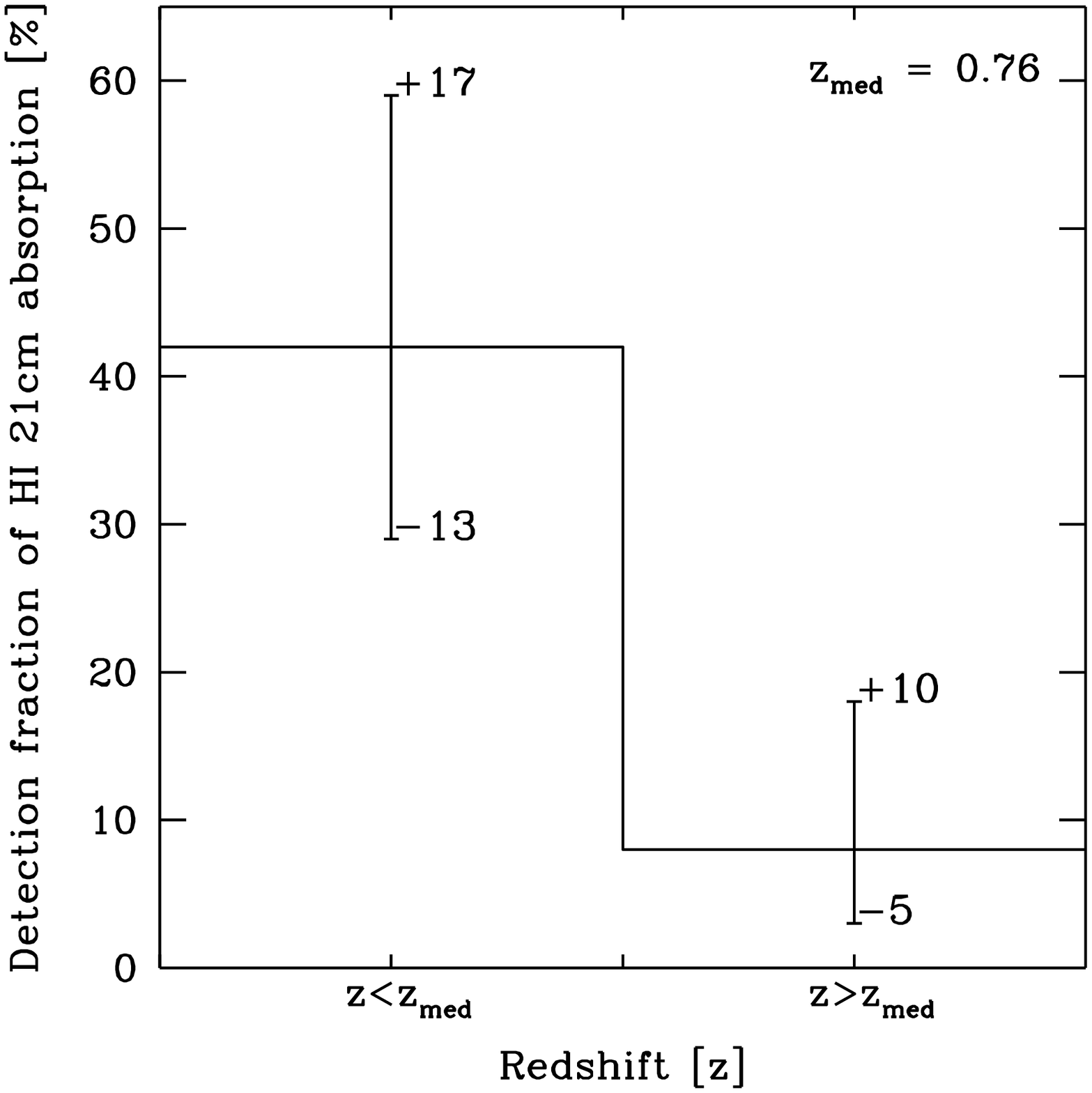}      \\
\end{tabular}
\end{figure*}

\begin{figure*}
\caption{[A]~Left panel: The rest-frame $1216$~\AA\ UV luminosity of 49 CJF sources 
of the sample plotted, in logarithmic units, versus the AGN redshift. The dotted horizontal 
line shows the median UV luminosity of the sample, $L_{\rm UV,med} = 10^{22.0}$~W~Hz$^{-1}$, while 
the dashed horizontal line shows the UV luminosity threshold of \citet{curran08} 
($L_{\rm UV} = 10^{23}$~W~Hz$^{-1}$).
[B]~Right panel: The rest-frame 1.4~GHz radio luminosity L$_{\rm 1.4\,GHz}$ of the 52~CJF sources of the sample 
plotted, in logarithmic units, versus the AGN redshift. The dashed horizontal line shows the 
median rest-frame 1.4~GHz radio luminosity of the sample, $L_{\rm 1.4\, GHz} = 10^{27.3}$~W~Hz$^{-1}$,
while the dashed vertical line shows the median redshift of the sample, $z_{\rm med} = 0.76$.
The 23 sources of the present sample and the 29 literature systems are shown by squares and triangles,
respectively, with filled and open symbols respectively indicating detections and non-detections of \hii\ 
absorption.
\label{fig:lum-z}}

\begin{tabular}{cc}
\includegraphics[scale=0.45]{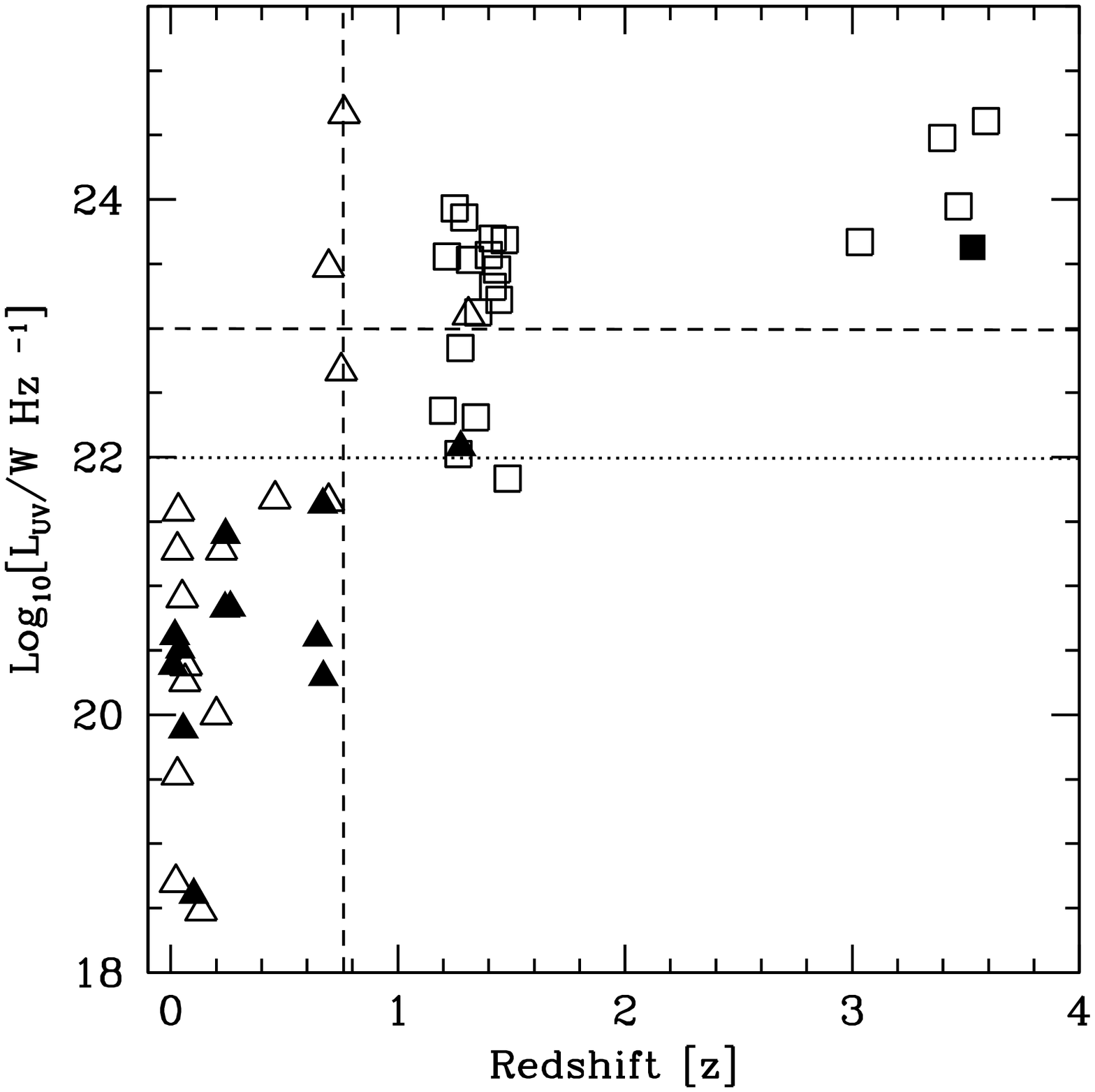}  &  \includegraphics[scale=0.45]{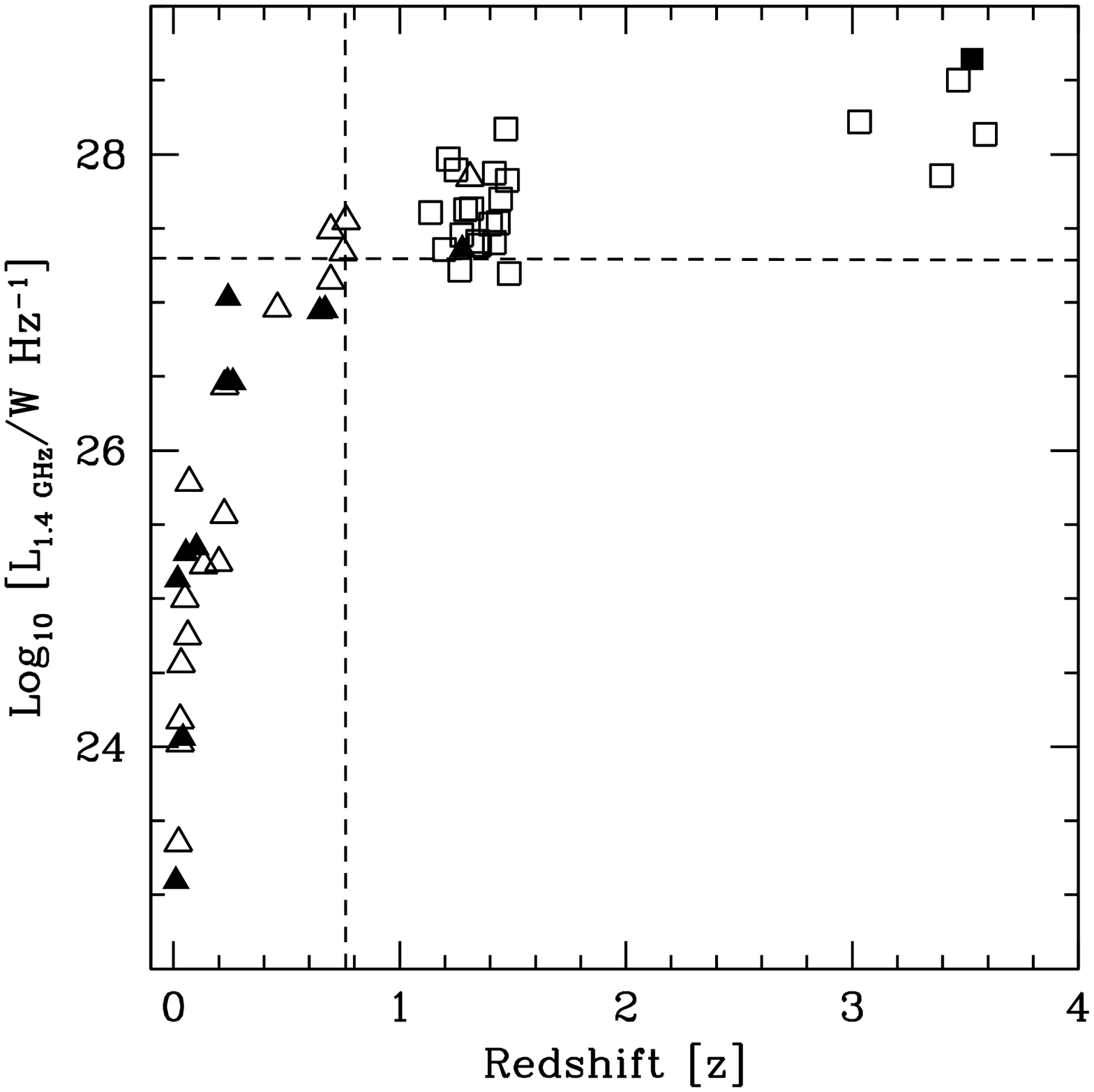}      \\
\end{tabular}
\end{figure*}

\begin{figure*}
\caption{The integrated \hii\ optical depth plotted against [A]~(Left panel)~the rest-frame 
$1216\;\AA$ UV luminosity, and [B]~(Right panel)~the rest-frame 1.4~GHz radio luminosity, with 
all quantities in logarithmic units. Again, the 23 AGNs of this paper and the 29 literature sources 
are shown as squares and triangles, respectively, while filled and open symbols indicate detections of 
\hii\ absorption and upper limits on the \hii\ optical depth. The dashed vertical lines in the left 
and right panels indicate, respectively, the median rest-frame $1216\;\AA$ UV luminosity 
($L_{\rm UV,med} = 10^{22.0}$~W~Hz$^{-1}$),
and the median rest-frame 1.4~GHz radio luminosity ($L_{\rm 1.4\, GHz} = 10^{27.3}$~W~Hz$^{-1}$).
\label{fig:tau-lum}}
\begin{tabular}{cc}
\includegraphics[scale=0.45]{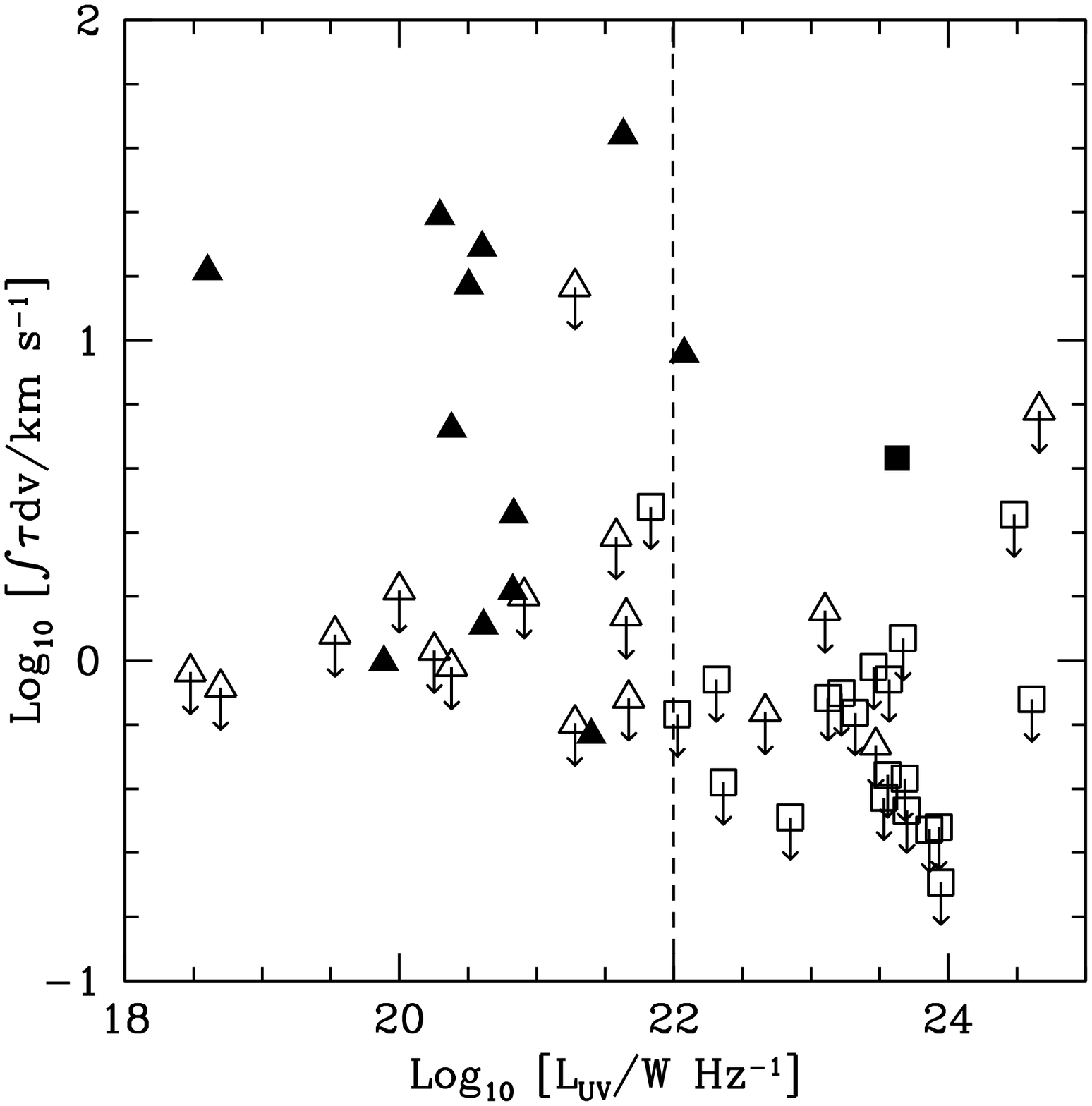}  &  \includegraphics[scale=0.45]{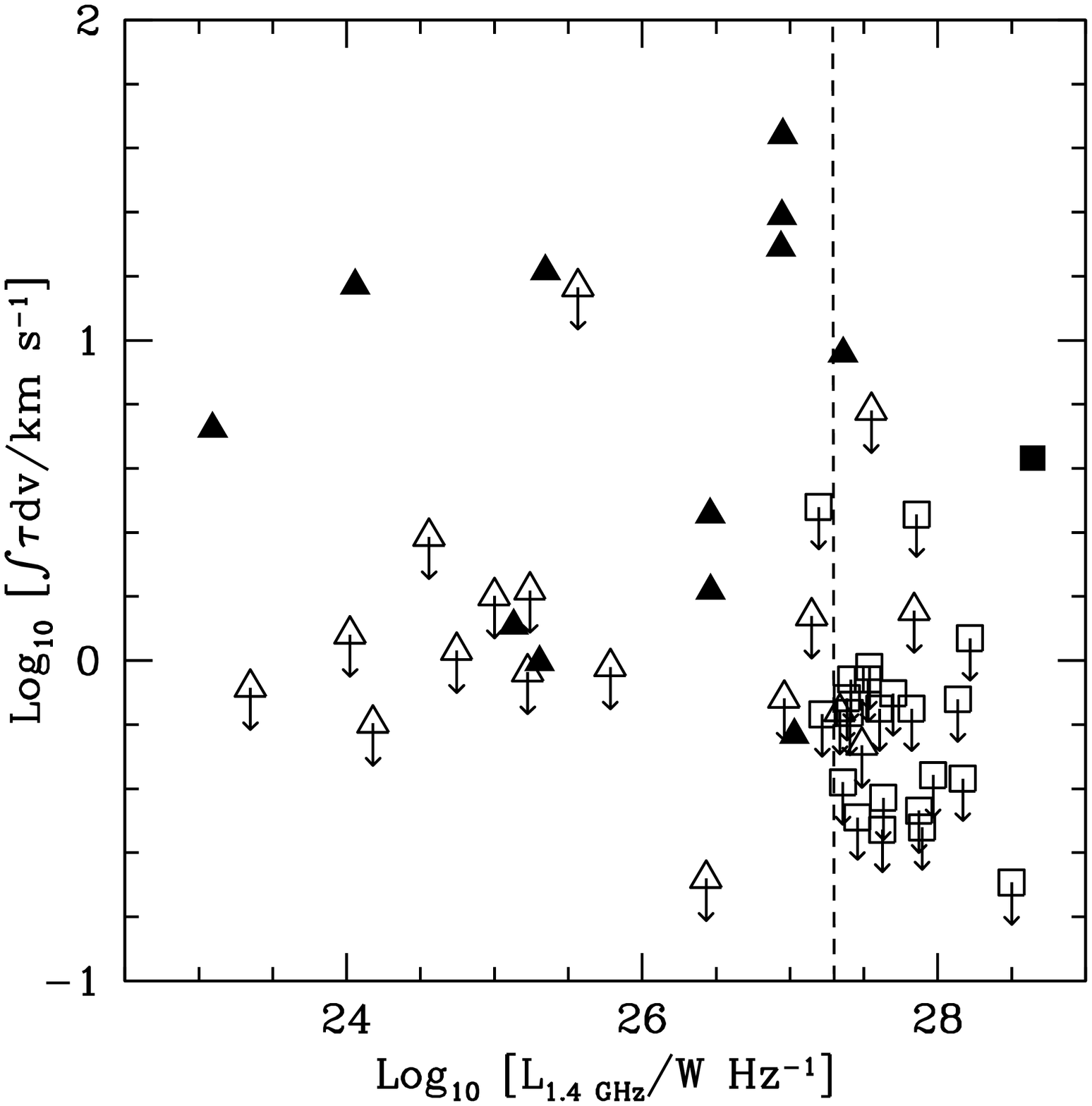}      \\
\end{tabular}
\end{figure*}

\begin{figure}
\caption{The integrated \hii\ optical depth plotted against the AGN low-frequency spectral
index $\alpha_{\rm 21\,cm}$, computed around the redshifted \hii\ line frequency. 
The 23 sources of this paper and the 29 literature sources are shown as squares and triangles,
respectively, with filled and open symbols representing \hii\ detections and upper limits, 
respectively. The dashed 
vertical line indicates the median low-frequency spectral index of the sample, 
$\alpha_{\rm 21\,cm,\;med} = -0.06$.
\label{fig:tau-alpha}}
\includegraphics[scale=0.45]{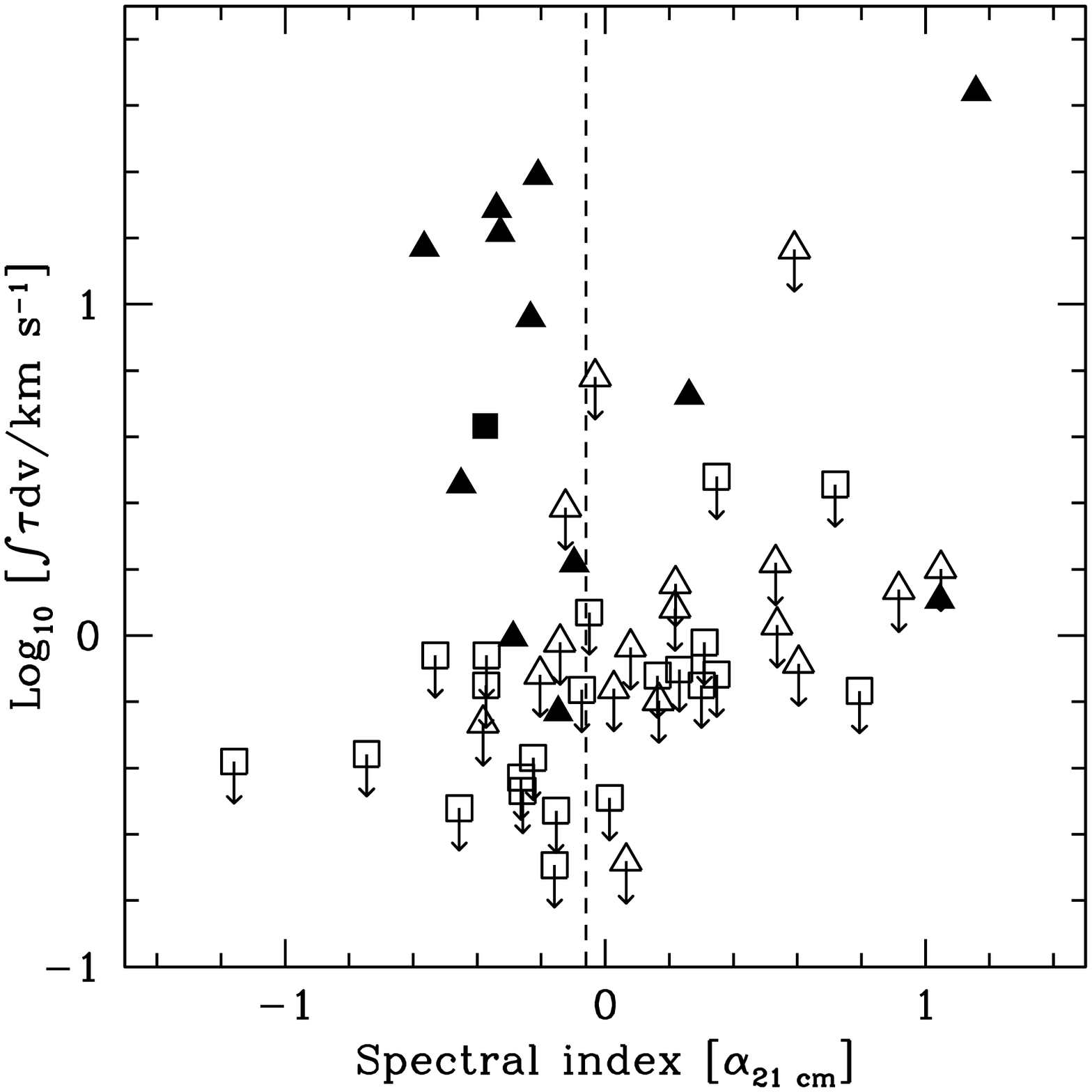}
\end{figure}

\begin{table*}
\caption{The 29 CJF sources with \hii\ absorption searches in the literature, 
in order of increasing redshift.
Note that L$'_{\rm UV}=$~Log[L$_{\rm UV}$/W~Hz$^{-1}$] and L$'_{\rm 1.4\,GHz} =$~Log[L$_{\rm 1.4\,GHz}$/W~Hz$^{-1}$].
\label{table:literature}}
\begin{center}
\begin{tabular}{|c|c|c|c|c|c|c|c|c|}
\hline
Source      & $z$ & $\int \tau dV $ & L$'_{\rm UV}$ & L$'_{\rm 1.4\,GHz}$ & $\alpha_{\rm 21\,cm}$ & Ref. \\ 
\hline
            &    &  \kms\     &      &       &     &      & & \\ \hline            
TXS\,1146+596  & 0.0108 & 5.30     & 20.30 &  23.09 & 0.26 & 1  \\ 
TXS\,0316+413  & 0.0176 & 1.28     & 20.61 &  25.13 & 1.05 & 2  \\  
B3\,0651+410   & 0.022  & $<0.82$  & 18.69 &  23.35 & 0.60 & 3  \\ 
TXS\,1101+384  & 0.030  & $<0.63$  & 21.28 &  24.18 & 0.17 & 4  \\ 
TXS\,1744+557  & 0.030  & $<1.2$   & 19.53 &  24.02 & 0.22 & 5  \\ 
TXS\,1652+398  & 0.034  & $<2.4$   & 21.58 &  24.56 &-0.12 & 4  \\ 
TXS\,1254+571  & 0.042  & 14.8     & 20.51 &  24.06 &-0.57 & 6  \\ 
TXS\,1807+698  & 0.051  & $<1.6$   & 20.91 &  25.00 & 1.05 & 4  \\ 
TXS\,0402+379  & 0.055  & 0.98     & 19.88 &  25.31 &-0.29 & 7  \\ 
TXS\,1144+352  & 0.063  & $<1.1$   & 20.26 &  24.75 & 0.54 & 5  \\ 
TXS\,2200+420  & 0.069  & $<0.95$  & 20.30 &  25.78 &-0.14 & 4  \\ 
TXS\,1946+708  & 0.101  & 16.46    & 18.60 &  25.35 &-0.33 & 8 \\ 
TXS\,0309+411  & 0.134  & $<0.92$  & 18.48 &  25.23 & 0.08 & 5  \\ 
IVS\,B1622+665 & 0.201  & $<1.7$   & 20.00 &  25.24 & 0.53 & 3  \\ 
S5\,1826+79    & 0.224  & $<14.6$  & 21.28 &  25.57 & 0.59 & 9  \\ 
TXS\,2021+614  & 0.227  & $<0.21$  & $-$   &  26.43 & 0.07 & 9  \\ 
TXS\,2352+495  & 0.238  & 1.65     & 20.84 &  26.46 &-0.09 & 9  \\ 
TXS\,0831+557  & 0.241  & 0.58     & 21.39 &  27.03 &-0.15 & 9  \\ 
TXS\,1943+546  & 0.263  & 2.86     & 20.85 &  26.46 &-0.45 & 9  \\ 
TXS\,1031+567  & 0.459  & $<0.76$  & 21.67 &  26.96 &-0.20 & 9  \\ 
TXS\,1355+441  & 0.646  & 19.4     & 20.60 &  26.94 &-0.34 & 9  \\ 
S4\,0108+38    & 0.669  & 43.77    & 21.63 &  26.95 & 1.16 & 10  \\ 
TXS\,1504+377  & 0.6715 & 24.4     & 20.30 &  26.95 &-0.21 & 10  \\ 
TXS\,0923+392  & 0.695  & $<0.54$  & 23.47 &  27.49 &-0.38 & 9  \\ 
S5\,0950+74    & 0.695  & $<1.37$  & 21.65 &  27.15 & 0.92 & 9  \\ 
TXS\,1642+690  & 0.751  & $<0.69$  & 22.66 &  27.34 & 0.03 & 9  \\  
S4\,1843+35    & 0.764  & $<6.0$   & 24.66 &  27.55 &-0.03 & 9  \\ 
TXS\,1543+480  & 1.277  & 9.1      & 22.08 &  27.36 &-0.23 & 11 \\ 
TXS\,0248+430  & 1.311  & $<1.4$   & 23.10 &  27.84 & 0.22 & 1  \\ 
  & & & & & & \\ \hline
\end{tabular}
\end{center}
\bigskip
References: (1)~\citealp{gupta06a}; (2)~\citealp{deyoung73}; (3)~\citealp{orienti06}; 
(4)~\citealp{vangorkom89}; (5)~\citealp{chandola13}; (6)~\citealp{dickey82a}; 
(7)~\citealp{morganti09}; (8)~\citealp{peck99}; (9)~\citealp{vermeulen03}; 
(10)~\citealp{carilli98}; (11)~\citealp{curran13}.
\end{table*}

\section{Discussion}
\label{sec:discuss}

\subsection{A uniformly-selected flat-spectrum sample}
\label{sec:full-sample}

As noted in the introduction, there are only four known associated \hii\ absorbers at 
$z > 1$, corresponding to a low apparent detection rate of \hii\ absorption. However, there are 
more than 30 detections of \hii\ absorption at low redshifts, $z < 1$ 
\citep[e.g.][]{vermeulen03,gupta06a,gereb15}, with a typical reported detection fraction 
of $\gtrsim 30$\% \citep[e.g.][]{pihlstrom03,vermeulen03}. This suggests that the neutral gas 
content in the AGN environment (or physical conditions therein) may evolve with redshift. 
However, the comparison between the \hii\ detection fractions has mostly been carried out with 
heterogeneous AGN samples, where it is difficult to separate redshift evolution in the 
AGN environment from differences in the AGN samples at different redshifts. For example, 
\citet{gupta06a} carried out a statistical analysis of 96~AGNs with associated \hii\ absorption 
studies, and found little evidence for redshift evolution in the detection rates of \hii\ 
absorption or the distribution of \hii\ optical depths. However, their sample was highly 
heterogeneous, consisting of 27 GPS sources, 35 compact steep spectrum sources, 13 flat-spectrum 
sources and 21 large radio galaxies; this makes it difficult to interpret their results. 

An alternative explanation of a possible lower strength of \hii\ absorption in high-$z$ 
AGNs arises from the fact that the high-$z$ AGN sample typically contains objects of 
higher rest frame luminosities, in both the radio and the UV wavebands. 
\citet{curran08} noted that a high AGN luminosity in either of these bands might result 
in a lower strength of \hii\ absorption. Specifically, it has long been known that proximity 
to a strong radio source can alter the hyperfine level populations, increasing the gas 
spin temperature, and thus lowering the \hii\ optical depth \citep{field59}. 
Conversely, \citet{curran08} pointed out that a high AGN luminosity at UV wavelengths, 
$\lesssim 1216$~\AA, might excite the electron in neutral hydrogen to higher energy 
levels (although this is unlikely to be an important effect, given the very short decay 
timescale to the ground state) or entirely ionize the gas, again lowering the strength 
of the \hii\ absorption. Both redshift evolution and dependence of local conditions on 
the AGN UV luminosity were found to be a viable explanations for the low observed strength 
of associated \hii\ absorption in high-$z$ AGN \citep{curran08,curran13}. Again, however, 
their sample was highly heterogeneous, consisting of all radio sources in the literature with 
searches for associated \hii\ absorption.

Our targets have been selected from the CJF sample as flat-spectrum sources lying in the 
redshift ranges $1.1 < z < 1.5$ and $3.0 < z < 3.6$. A number of CJF sources, especially 
at lower redshifts $z < 0.7$, have \hii\ absorption studies in the literature 
\citep[e.g.][]{vangorkom89,orienti06,gupta06a,vermeulen03,curran13}. We have included 
these literature sources in our full sample in order to investigate the dependence of the \hii\ 
absorption strength on both redshift and AGN luminosity in a uniformly-selected sample. 
The literature sources are listed (ordered by increasing redshift) in Table~\ref{table:literature}, whose 
columns are (1)~the AGN name, (2)~the AGN redshift, (3)~the integrated \hii\ optical 
depth or, for \hii\ non-detections, the $3\sigma$ limit to the integrated \hii\ optical depth, 
in \kms\ (assuming a line FWHM of 100~\kms, (4)~the AGN luminosity at a 
rest-frame wavelength of $1216\;\AA$ (see Section~\ref{sec:results} for details), (5)~the AGN 
luminosity at a rest-frame frequency of 1.4~GHz, (6)~the AGN spectral index around the redshifted
\hii\ line frequency, $\alpha_{\rm 21\,cm}$, and (7)~the literature reference to the \hii\ 
absorption study. There are 29~sources in the 
literature sample, with 12 detections of \hii\ absorption (11 at $z < 0.7$, and one at $z=1.277$),
and seventeen non-detections, sixteen of which are at $z < 0.8$. For one source, TXS\,2021+6134,
the luminosity is only known at a single optical waveband, far from the redshifted $1216\,\AA$ 
wavelength; its rest-frame $1216\,\AA$ UV luminosity is hence not listed in the table.

Our full sample thus consists of 52~sources selected from the CJF sample with associated \hii\ 
absorption studies, 23 from our observations and 29~from the literature, with 13 detections of 
\hii\ absorption (including 
our tentative detection towards TXS\,0604+728) and 39~non-detections. This is the largest 
uniformly-selected sample of AGNs with redshifted \hii\ absorption studies; the median redshift 
of the sample is $z_{\rm med} = 0.76$.  In the following sections, we will examine the dependence 
of the \hii\ detection fractions and the distribution of \hii\ optical depths on redshift, radio 
spectral index, and AGN radio and UV luminosity.

\subsection{Redshift evolution}
\label{sec:redshift}

The left panel of Fig.~\ref{fig:tau} plots the integrated \hii\ optical depth (in logarithmic 
units) versus redshift for our full sample of 52~sources. The high sensitivity of our 
GMRT observations imply that the $3\sigma$ limits on the integrated \hii\ optical depths 
of our non-detections are in almost all cases (especially at redshifts $1.1. < z < 1.5$) 
sufficiently stringent to rule out \hii\ opacities comparable to those of the low-$z$ 
detections of \hii\ absorption. It is clear from the figure that the detections of \hii\ absorption 
in this uniformly-selected flat-spectrum sample are concentrated in the redshift range
$z < 0.7$, with only one detection at $z = 1.277$ \citep[][]{curran13}, and a tentative 
detection at $z \approx 3.530$ (this work). On dividing the sample at the median redshift, 
$z_{\rm med} = 0.76$, the low-$z$ sample has 11 detections and 15 non-detections, i.e. 
a detection fraction of $42^{+17}_{-13}$\%, while the high-$z$ sample has 2 detections 
(one of which is tentative) and 24 non-detections, for a detection fraction of $8^{+10}_{-5}$\% 
\citep[where the errors are from small-number Poisson statistics; e.g.][]{gehrels86}. 
Thus, although the detection fraction appears to be higher at low redshifts, the difference 
in detection fractions in the high-$z$ and low-$z$ samples has only $\approx 2.1\sigma$ 
significance.

However, it also appears from Fig.~\ref{fig:tau}[A] that the measured \hii\ optical 
depths for the low-$z$ sample are higher than the typical upper limits to the \hii\ optical 
depths of the high-$z$ sample. We hence used the Astronomical Survival analysis {\sc asurv}
package \citep{isobe86} to investigate whether there is indeed a difference in the distributions
of the \hii\ optical depths of the low-$z$ and high-$z$ samples, taking into account
the fact that some of the measurements are censored, i.e. are limits to the \hii\
optical depths. A Peto-Prentice generalized Wilcoxon two-sample test (for censored 
data) finds weak evidence that the \hii\ optical depths of the low-$z$ and high-$z$ 
samples are drawn from different distributions: the null hypothesis that they are 
drawn from the same sample is rejected at $2.7\sigma$ significance. Note that this 
result assumes the \hii\ absorption tentatively detected at $z=3.530$ towards 
TXS\,0604+728 is a real absorption feature. If this system (for which the \hii\ absorption 
remains to be confirmed) is excluded from the sample, the null hypothesis that the 
two samples are drawn from the same distribution is ruled out at $\approx 3.1\sigma$ 
significance in the Peto-Prentice test. While observations of a larger sample are 
needed to draw statistically reliable conclusions, we conclude that the present data 
show tentative evidence for redshift evolution in the strength of \hii\ absorption 
in AGN environments. 

In passing, we note that, should the evidence for redshift evolution in the
strength of associated \hii\ absorption be confirmed, this could imply either a lower 
typical \hi\ column density or a lower cold gas fraction (i.e. a higher spin temperature) 
in the environments of high-$z$ flat-spectrum AGNs. Unfortunately, this issue has often been ignored 
in studies of associated \hii\ absorption, where one does not have an independent 
estimate of the \hi\ column density \citep[unlike in the case of intervening \hii\ 
absorbers towards quasars, where one can usually determine the \hi\ column density 
from the damped Lyman-$\alpha$ absorption line; e.g.][]{wolfe05}. Most associated 
\hii\ absorption studies in the literature assume a single spin temperature (usually, 
100 or 1000~K) for the neutral gas in all AGNs and use this to convert the \hii\ 
optical depth into an \hi\ column density, and then discuss the evolution of the 
\hi\ column density with redshift. We emphasize that it is very difficult to justify 
this assumption of a uniform spin temperature in AGN environments. Indeed, it is 
clear from \hii\ absorption studies of intervening damped Lyman-$\alpha$ systems (DLAs) 
that absorption-selected galaxies at high redshifts typically have high spin temperatures,
indicating larger fractions of the warm neutral medium 
\citep[e.g.][]{chengalur00,kanekar03,kanekar14}; this may well also be the case 
for AGN environments. It would hence be more appropriate in studies of associated 
\hii\ absorption to consider the dependence of the \hii\ optical depths, rather than
the \hi\ column density, on redshift, AGN luminosity, etc..

\subsection{Dependence on the radio spectral index}

A possible cause for the weakness of \hii\ absorption in the high-$z$ AGN sample is that the 
radio emission of the high-$z$ AGNs might be dominated by extended structure. While the 
targets have been chosen from the CJF sample, and are hence flat-spectrum sources 
\citep[with $\alpha \geq -0.5$;][]{taylor96}, this spectral index is over the frequency 
range $1400-4850$~MHz. At the lower frequencies of the redshifted \hii\ line, it is possible
that the flat- or inverted-spectrum AGN core contributes a smaller fraction of the radio 
emission than the steep-spectrum extended structure. If so, the covering factor of 
neutral gas in the AGN environment could be low for high-$z$ sources, thus reducing the 
apparent \hii\ optical depth. This might be tested by VLBI imaging of the radio 
continuum at or near the redshifted \hii\ line frequency 
\citep[i.e. at frequencies $< 1$~GHz; e.g.][]{kanekar09a}. Unfortunately, such 
low-frequency VLBI studies are not available for most of our target sources, especially the 
ones at high redshifts, $z > 1$. We will hence use the AGN spectral index {\it around the redshifted 
\hii\ line frequency} $\alpha_{\rm 21\,cm}$ as a proxy to estimate the compactness or lack thereof 
of the low-frequency radio emission: a flat spectrum would indicate core-dominated emission and 
a steep spectrum, extended structure. Fig.~\ref{fig:tau-alpha} plots the integrated \hii\ optical 
depth versus $\alpha_{\rm 21\,cm}$; the dashed vertical line indicates the median low-frequency 
spectral index, $\alpha_{\rm 21\,cm,\; med} = -0.06$. The figure shows that there are indeed 
a few sources with low-frequency spectral indices $< -0.5$, i.e. steeper than the high-frequency
spectral indices. However, we note that the median low-frequency spectral index remains close 
to 0, suggesting that the effect is not a strong one; indeed, there are only four sources
with $\alpha_{\rm 21\,cm} < -0.5$. A Peto-Prentice two-sample test finds that the low-$\alpha_{\rm 21\,cm}$ 
and high-$\alpha_{\rm 21\,cm}$ sub-samples (separated at $\alpha_{\rm 21\,cm,\; med} = -0.06$) 
are consistent with being drawn from the same distribution (within $1.8\sigma$ significance).
Excluding the tentative \hii\ detection towards TXS\,0604+728 has only a marginal effect 
on the results: the sub-samples are then consistent with being drawn from the same distribution
within $1.6\sigma$ significance. We conclude that there is no significant evidence that 
the apparent weakness in the \hii\ absorption in the high-$z$ AGN sample arises due to the 
low-frequency radio emission being dominated by extended, steep-spectrum structure, and hence, 
a low AGN covering factor.

\subsection{Dependence on the AGN luminosity}

As noted above, the lower strength of \hii\ absorption in high-redshift AGNs
could also arise if the high-$z$ AGN sample is systematically biased towards objects 
with high UV or radio luminosities \citep[][]{curran08}. Fig.~\ref{fig:lum-z} tests 
this hypothesis by plotting (left panel) the rest-frame $1216\;\AA$ UV luminosities and 
(right panel) the rest-frame 1.4~GHz radio luminosities versus redshift (note that only 
49 sources of the 52 are included in the left panel, as the remaining three do not 
have reliable estimates of their UV luminosity). It is apparent from the figure that AGNs of 
the high-$z$ sample have significantly higher rest-frame $1216\;\AA$ UV and 1.4~GHz radio 
luminosities than those of the low-$z$ sample. Again dividing the two samples at the median 
redshift $z_{\rm med} = 0.76$, a Peto-Prentice two-sample test finds that the null hypothesis 
that the rest-frame UV luminosities of the two samples are drawn from the same distribution 
is ruled out at $6.4\sigma$ significance. Similarly, the null hypothesis that the rest-frame 
1.4~GHz luminosities of the low-$z$ and high-$z$ samples are drawn from the same distribution 
is ruled out at $6.7\sigma$ significance. We conclude that our present flat-spectrum sample 
contains a strong bias towards higher rest-frame UV and 1.4~GHz luminosities at high redshifts.
Similarly, the rest-frame UV and rest-frame 1.4~GHz luminosities are also strongly correlated
for the sources of our sample.

We hence considered the possibility that the apparent redshift evolution in 
the strength of the \hii\ absorption might arise due to the above luminosity bias.
The two panels of Fig.~\ref{fig:tau-lum} plot the integrated \hii\ optical depths 
(in logarithmic units) versus (left panel) the rest-frame $1216\;\AA$ UV luminosity,
and (right panel) the rest-frame 1.4~GHz radio luminosity, both in logarithmic units. 
The median $1216\;\AA$ UV luminosity of our sample is ${\rm Log}[L_{\rm UV,med}/{\rm W Hz}^{-1}] 
= 22.0$, while the median 1.4~GHz luminosity of our sample is ${\rm Log}[L_{\rm 1.4\,GHz,med}/{\rm W Hz}^{-1}] 
= 27.3$. Dividing the low-luminosity and high-luminosity samples at the median luminosity 
for each case, a Peto-Prentice two-sample test finds that the hypothesis that the 
\hii\ optical depth distributions of the low- and high-luminosity samples are drawn
from the same distribution is ruled out at $3.4\sigma$ significance for the rest-frame 
$1216\,\AA$ UV luminosity and at $3.1\sigma$ significance for the rest-frame 
1.4~GHz radio luminosity. As above, if we exclude the tentative detection of \hii\ 
absorption towards TXS0604+728 from the sample, the above hypotheses are ruled 
out at $3.7\sigma$ and $3.4\sigma$ significance, respectively. 

We thus find that the strength of the \hii\ absorption has a statistically significant 
dependence on redshift, rest-frame 1.4~GHz radio luminosity, and rest-frame $1216\,\AA$ UV 
luminosity, with weaker absorption being obtained at high redshifts and high AGN luminosities. 
Unfortunately, the luminosity bias in the sample, with the higher-luminosity AGNs located 
at higher redshifts, implies that the present sample does not allow us to distinguish 
between the three possibilities, and identify the primary cause, if any, of the decline 
in the strength of the \hii\ absorption in the high-$z$, high-luminosity sample. Observations 
of either a low-luminosity AGN sample at high redshifts or a high-luminosity AGN sample 
at low redshifts would be needed to break the degeneracy.

\begin{figure}
\caption{The AGN low-frequency spectral index $\alpha_{\rm 21\,cm}$, computed around the 
redshifted \hii\ line frequency, plotted versus redshift. The 23 sources of this paper and 
the 29 literature sources are shown as squares and triangles, respectively. The dashed 
vertical line indicates the median redshift, $z = 0.76$.
\label{fig:alpha-z}}
\includegraphics[scale=0.45]{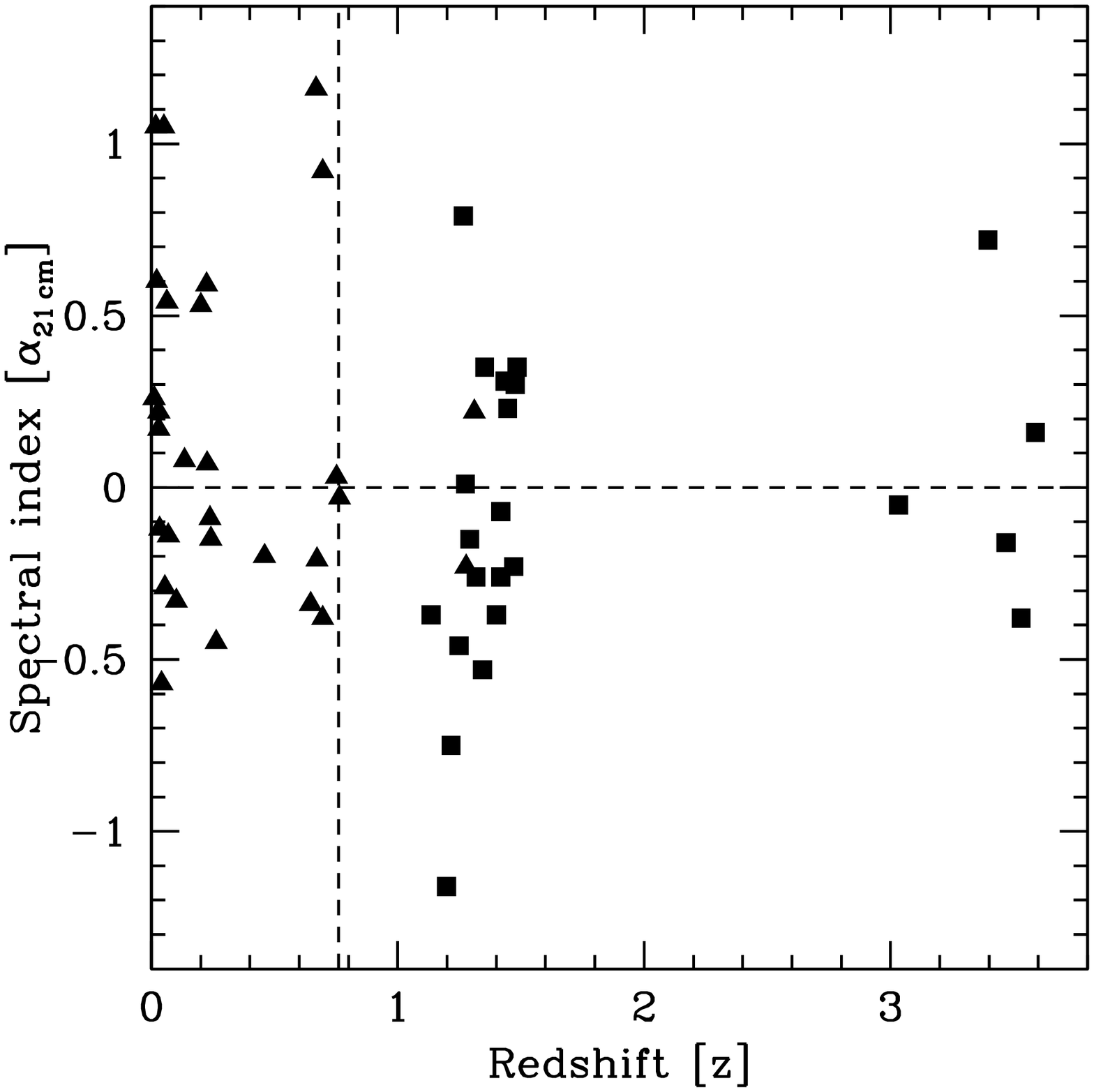}
\end{figure}

\subsection{Covering factor issues}

As mentioned earlier, there is no evidence that the non-detections of \hii\ absorption in
the high-$z$ CJF sample arise due to a low covering factor. However, a critical assumption in the 
interpretation of such associated \hii\ absorption studies, both in the present analysis and in 
the literature, is that the gas covering factor is unity for all systems, i.e. that the radio 
emission is entirely covered by any foreground gas. This is manifestly a poor assumption for 
certain classes of radio sources: for example, in Fanaroff-Riley-II galaxies, the radio emission 
is usually dominated by the large-scale lobes \citep[e.g.][]{odea98}. Even for compact 
steep spectrum sources, it is possible that the neutral gas lies close to the AGN core, and 
hence does not cover a significant fraction of the somewhat more extended steep spectrum source 
components. In this section, we hence consider the possibility that the AGN covering factor 
might be significantly different from unity, and its implications for our results.

First of all, it is clear that the results suggesting redshift evolution or AGN luminosity 
dependence are unchanged if {\it all} AGNs of the sample have low covering factors, provided 
the covering factors are the same. Further, the results of Section~\ref{sec:redshift} suggest 
that the \hii\ absorption is stronger for low-$z$ AGNs, even when the covering factor is 
assumed to be unity. If low-$z$ AGNs have low covering factors, $f \ll 1$, we would have 
under-estimated the strength of their \hii\ absorption by assuming $f=1$. In other words, 
a covering factor lower than unity for the low-$z$ population can only increase the 
significance of the apparent redshift evolution. We hence consider three possibilities here, 
(1)~AGNs at all redshifts have random covering factors, between 0 and unity, 
(2)~low-$z$ AGNs have high covering factors, $\approx 1$, while high-$z$ AGNs have 
random covering factors, uniformly distributed between 0 and unity, and (3)~low-$z$ AGNs 
have high covering factors, $f \approx 1$, while high-$z$ AGNs have low covering factors, 
$f \approx 0.1$. 

For the first scenario, of all AGNs having random covering factors, the Peto-Prentice test finds that 
the null hypothesis that the low-$z$ and high-$z$ AGNs are drawn from the same population is rejected
at $\approx 2.9\sigma$ and $\approx 3.2 \sigma$ significance (depending, respectively, on whether 
we include or exclude the tentative detection of \hii\ absorption towards TXS\,0604+728). These
are very similar to the results on assuming a uniform covering factor of unity. Similarly, 
the null hypothesis that the low- and high-UV luminosity AGNs are drawn from the same 
distribution is rejected at $\approx 3.1\sigma$ (retaining TXS\,0604+728) and $\approx 3.4\sigma$ 
(excluding TXS\,0604+728) significance, while the hypothesis that the low- and high-radio 
luminosity AGNs are drawn from the same distribution is rejected at $\approx 3.0\sigma$ (retaining
TXS\,0604+728) and $\approx 3.3\sigma$ (excluding TXS\,0604+728) significance. In other words,
the results do not significantly change if one assumes that the AGNs of the sample have random covering 
factors, uniformly distributed between 0 and 1, instead of a single covering factor of unity.

For the second scenario, of the low-$z$ AGNs having a covering factor of unity and the high-$z$ ($z \geq 1$)
AGNs having random covering factors, we find that the null hypotheses that the sub-samples are 
drawn from the same underlying distribution are rejected at $\approx 2.4\sigma$ and $\approx 2.7\sigma$ 
significance (with and without TXS\,0604+728, respectively, for the low-$z$ and high-$z$ sub-samples), 
at $\approx 2.7\sigma$ and $\approx 3.1\sigma$ significance (with and without TXS\,0604+728, respectively, 
for the low- and high-UV luminosity sub-samples), and at $\approx 2.6\sigma$ and $\approx 3.0\sigma$ 
significance (with and without TXS\,0604+728, respectively, for the low- and high-radio luminosity 
sub-samples. While the statistical significances at which the null hypothesis is rejected are 
somewhat lower than in the case of unity covering factors, the results are again not significantly 
different.

Finally, in the third scenario, wherein low-$z$ AGNs have a covering factor of unity and high-$z$ ($z \geq 1$)
AGNs have low covering factors ($f = 0.1$), we find that the null hypotheses that the sub-samples
are drawn from the same distribution are rejected at far lower significance in all cases, $\approx 1.2\sigma$
and $\approx 1.7\sigma$ significance for the redshift sub-samples, $\approx 1.7\sigma$ and $\approx 2.2\sigma$
significance for the UV luminosity sub-samples, and $\approx 1.6\sigma$ and $\approx 2.1\sigma$ significance
for the radio luminosity sub-samples (where, in all cases, the two values correspond to retaining and 
excluding TXS\,0604+728). Thus, if high-$z$ AGNs indeed have systematically lower covering factors
than low-$z$ systems, the present evidence for redshift evolution (or even for a luminosity dependence 
in the \hii\ absorption strength) is not statistically significant.

To test whether the high-$z$ AGNs are indeed likely to have systematically lower covering factors than
the low-$z$ systems, we inspected the distribution of the radio spectral indices of the AGN sub-samples 
as a function of redshift. If the high-$z$ AGNs have significantly steeper spectra at low frequencies,
the radio emission of these systems is likely to arise from extended structure, rather than the core; 
if so, the high-$z$ AGNs might have a systematically lower covering factor. Figure~\ref{fig:alpha-z} 
plots the low-frequency radio spectral index ($\alpha_{\rm 21\,cm}$) against redshift for our full 
sample of 52 AGNs; there is no obvious difference between the spectral indices of the low- and high-$z$ 
sub-samples (separated by the dashed vertical line, at $z=0.76$). A Gehan Wilcoxon two-sample test rules 
out the null hypothesis that the two sub-samples are drawn from the same distribution at $\approx 1.4\sigma$
significance, implying that the sub-samples are consistent with being drawn from the same distribution. 
Given that there appears to be no significant difference between their radio spectral indices, it 
appears unlikely that the high-$z$ AGNs of our sample indeed have systematically lower covering factors than 
their low-$z$ counterparts.

The results concerning the redshift and luminosity dependence of the strength of the \hii\ absorption 
thus appear to be robust if the covering factors are assumed to be random, either at all redshifts or 
only at high redshifts. It is only if the covering factors are assumed to be high for low-$z$ AGNs and 
systematically low for high-$z$ AGNs that the null hypothesis that the sub-samples are drawn from the 
same population cannot be ruled out at high (i.e. $\approx 3\sigma$) significance. This scenario 
appears unlikely, as the high-$z$ and low-$z$ AGN samples have similar spectral index distributions. We
hence conclude that our results do not appear to strongly depend on the assumption that the AGN covering 
factor is unity.

\subsection{The \hii\ absorber at $z = 3.530$ towards TXS\,0604+728}

If confirmed, the absorption feature towards TXS\,0604+728 would be at the highest redshift 
at which \hii\ absorption has ever been discovered, surpassing the two absorbers at 
$z \approx 3.39$ towards TXS\,0902+343 \citep[][]{uson91} and PKS\,0201+113 \citep[][]{kanekar07}. 
The putative \hii\ absorption is extremely wide, with a full width between 20\% points of 
$\approx 850$~\kms. For comparison, the associated \hii\ absorbers in the samples of 
\citet{gereb15} or \citet{vermeulen03} all have velocity widths (between 20\% points) 
$\leq 825$~\kms. The absorption appears symmetric about the source redshift, but the 
signal-to-noise ratio of the present spectrum is not sufficient to 
derive detailed kinematic information.

TXS\,0604+728 has a complicated core-jet structure on angular scales ranging from milli-arcsecs 
to $\approx 10$~arcsecs \citep[e.g.][]{taylor94,taylor96,britzen07}. It has a spectral 
index of $\alpha = -0.35$ ($S_\nu \propto \nu^\alpha$) between 1.4~GHz and 4.8~GHz 
\citep[][]{taylor96}, steepening slightly to $\alpha \approx -0.44$ between 325~GHz and 
1.4~GHz \citep[using 1420~MHz and 327~MHz flux densities from the NVSS and the 
Westerbork Northern Sky Survey, respectively;][]{condon98,rengelink97}. The wide absorption
profile could arise due to absorption against different radio source components in the core 
and the jets.

Finally, TXS\,0604+728 has a high estimated UV luminosity, $\approx 4.2 \times 10^{23}$~W~Hz$^{-1}$,
using a template to extrapolate from the measured luminosities in optical wavebands. If the
reality of the absorption feature of Fig.~\ref{fig:0604} is confirmed, it would be the 
first case of a detection of \hii\ absorption in an AGN with UV luminosity $\gtrsim 
10^{23}$~W~Hz$^{-1}$. This UV luminosity has been suggested as the threshold above
which the \hi\ in the AGN environment is either ionized or excited to higher energy 
states \citep[see discussion below;][]{curran11}; the latter authors argue that \hi\ 
absorption should hence not be detectable in AGNs with UV luminosities above this 
threshold. While there is little evidence in support of this argument (as discussed 
above, the lack of \hii\ absorption in present high-$z$, high-luminosity samples may 
arise either from redshift evolution, or due to a UV or a radio luminosity threshold, to 
mention just three possibilities), it also only applies to \hi\ in the vicinity of the 
AGN. In the case of AGNs with extended radio continuum, such as TXS\,0604+728, more distant 
neutral gas could well give rise to \hii\ absorption.

\section{Summary}
In summary, we have used the GMRT to carry out a search for redshifted \hii\ absorption
from neutral gas associated with 24 flat-spectrum AGNs, at redshifts $1.1 < z < 1.5$ and $3.0 < z < 3.6$,
and selected from the Caltech-Jodrell Flat-spectrum sample. We obtained a single tentative detection of 
\hii\ absorption, at $z = 3.530$ towards TXS\,0604+728, and 22 non-detections of \hii\ absorption, 
with stringent constraints on the \hii\ optical depth; the data on the last target was not usable due to RFI. 
Including data from the literature, 52 CJF sources have now been searched for associated \hii\ 
absorption, with a median sample redshift of $z_{\rm med} = 0.76$. We used two-sample tests to find 
$\approx 3\sigma$ evidence that the \hii\ absorption is stronger at low redshifts, and towards AGNs with 
low UV/radio luminosities. The results are robust to different assumed covering factors, except for the 
case that the AGN covering factors are assumed to be systematically low in the high-$z$ sub-sample. 
Using the radio spectral index as a measure of the AGN compactness, we find no evidence that the high-$z$ 
AGNs are systematically less compact than the low-$z$ AGNs; it hence appears unlikely that the high-$z$ 
AGN sub-sample indeed have systematically and significantly lower covering factors than the low-$z$ 
sub-sample. This is the first time that statistically significant evidence has been obtained 
for a redshift or luminosity dependence of the strength of \hii\ absorption in a uniformly selected 
sample. Unfortunately, the luminosity bias of the sample, with high-luminosity systems predominantly 
arising at high redshifts, implies that it is currently not possible to separate redshift evolution 
and AGN luminosity as the causes for the weakness of the \hii\ absorption in high-$z$, high-luminosity
active galactic nuclei.

\section*{Acknowledgements}

We thank Jayaram N. Chengalur for comments on an earlier version of this paper. We also thank the 
staff of the GMRT who have made these observations possible. The GMRT is run 
by the National Centre for Radio Astrophysics of the Tata Institute of Fundamental Research.
SK and NK acknowledge support from, respectively, the NCRA-TIFR Visiting Students' Research 
Programme and the Department of Science and Technology through a Swarnajayanti Fellowship
(DST/SJF/PSA-01/2012-13).

\bibliographystyle{mn2e}
\bibliography{ms}

\label{lastpage}
\end{document}